\documentclass[11pt,letterpaper]{article}

\usepackage[backend=biber,
  style=alphabetic,
  sorting=nyt,
  natbib=true,
  minalphanames=3,
  maxcitenames=99,
maxbibnames=99]{biblatex}

\usepackage{amsmath,amsthm,thmtools,amsfonts,amssymb}
\usepackage{mathrsfs}
\usepackage{mathtools}
\usepackage{physics}
\usepackage{braket}
\usepackage{array}
\allowdisplaybreaks

\usepackage[margin=1in]{geometry}
\usepackage{xspace}
\usepackage{xcolor}
\usepackage{subcaption}
\usepackage{microtype}
\usepackage{csquotes}
\usepackage{dsfont}

\usepackage[inline]{enumitem}
\setlist[itemize]{label=--}
\setlist[enumerate]{label=(\arabic*),labelindent=0.5\parindent,leftmargin=*}

\definecolor{citecolor}{HTML}{000080}
\definecolor{urlcolor}{HTML}{000080}
\usepackage{hyperref}
\hypersetup{
  colorlinks=true,
  linkcolor=citecolor,
  citecolor=citecolor,
  filecolor=urlcolor,
  urlcolor=urlcolor,
}

\usepackage[capitalize]{cleveref}
\AtBeginDocument{}
\crefname{step}{step}{steps}
\Crefname{step}{Step}{Steps}
\Crefname{equation}{Eq.}{Eqs.}

\usepackage[skins,many]{tcolorbox}

\makeatletter
\newcounter{HALG@line}
\renewcommand{\theHALG@line}{\thealgorithm.\arabic{ALG@line}}
\makeatother
\newtheorem{theorem}{Theorem}
\newtheorem{lemma}[theorem]{Lemma}

\newtheorem{corollary}[theorem]{Corollary}

\theoremstyle{definition}
\newtheorem{definition}[theorem]{Definition}

\newcommand{\prefix}[1]{\mathsf{P}#1}
\newcommand{\suffix}[1]{\mathsf{S}#1}
\newcommand{\out}{\mathsf{out}}
\newcommand{\Z}{\mathbb{Z}}
\newcommand{\N}{\mathbb{N}}

\newcommand{\calC}{\mathcal{C}}
\newcommand{\calF}{\mathcal{F}}
\newcommand{\calR}{\mathcal{R}}

\DeclareMathOperator{\polylog}{polylog}

\newcommand{\lift}[1]{{#1}^{\star}}
\input{figures/figures}

\begin{document}

\begin{flushleft}
  \huge\bf\boldmath
  Self-stabilising approximate clock synchronisation requires linear time
\end{flushleft}
\smallskip

\newcommand{\myemail}[1]{\,$\cdot$\, {\small #1}}
\newcommand{\myaff}[1]{\,$\cdot$\, {\small #1}\par\medskip}
\newenvironment{myabstract}{\list{}{\listparindent 1.5em \itemindent \listparindent \leftmargin 0cm \rightmargin 0cm \parsep 0pt}
\item\relax}{\endlist}
\newenvironment{mycover}{\list{}{\listparindent 0pt \itemindent \listparindent \leftmargin 0cm \rightmargin 1.5cm \parsep 0pt} \raggedright
\item\relax}{\endlist}

\begin{mycover}

  \textbf{Joel Rybicki}  \myemail{joel.rybicki@hu-berlin.de}
  \myaff{Humboldt-Universität zu Berlin, Germany}

  \bigskip
\end{mycover}

\thispagestyle{empty}

\medskip
\begin{myabstract}
  \noindent\textbf{Abstract.} Synchronising the clocks of processes is a fundamental problem in distributed computing.
  We consider message-passing systems in which processes start in an arbitrary initial state (self-stabilisation) and up to $f$ processes may be non-transiently faulty. It is well known that synchronising clocks \emph{exactly} is at least as hard as consensus and requires $\Theta(f)$ time. In contrast, without self-stabilisation, clocks can be synchronised \emph{approximately} much faster, in time independent of the number of non-transient faults [Srikanth and Toueg, JACM 1987; Welch and Lynch, Inf. Comput.~1988].

  In this work,
  we show that deterministic self-stabilising approximate clock synchronisation tasks require $\Omega(f)$ time to stabilise.
  This gives the first non-trivial time lower bound for these tasks, whose time complexity has remained open for over two decades [Dolev and Welch, JACM 2004; Dolev, F\"ugger, Schmid and Lenzen, JACM 2014].
  In particular, this answers the open question whether self-stabilising approximate clock synchronisation tasks are as hard as consensus or as easy as approximate agreement [Lenzen and Rybicki, JACM 2019].

  The results show that guaranteeing \emph{both} transient and non-transient fault-tolerance simultaneously is expensive even for approximate synchronisation tasks.
  The lower bounds hold in a very strong model: the faulty processes can only crash, the local clocks of processes have no clock drift, and the message delay is bounded and known to all processes.
  The proof is based on a combinatorial topological argument showing that no fast finite-state protocol can ensure that the clocks behave consistently across all executions.
  To our knowledge, this is the first  use of topological arguments to prove time lower bounds in the self-stabilising setting.
\end{myabstract}

\newpage

\tableofcontents

\thispagestyle{empty}
\setcounter{page}{0}
\newpage
\section{Introduction}

Keeping accurate time reliably is a fundamental problem in distributed systems~\cite{lamport1978time}. Many distributed algorithms assume some degree of global synchronisation among processes~\cite{lynch1996distributed}.
However, establishing and maintaining such synchronisation  can in itself be a highly challenging task: it is not feasible to predict all possible failure and attack patterns that may occur.
For example, unexpected \emph{transient} failures can corrupt the \emph{state} of the system, whereas
\emph{non-transient} failures caused by malfunctioning hardware or compromised software can alter the \emph{behaviour} of processes.

Distributed algorithms that tolerate arbitrary non-transient failures are called \emph{Byzantine-tolerant} \cite{pease1980reaching,lamport82byzantine}.
Algorithms that recover from arbitrary transient failures are called \emph{self-stabilising}~\cite{dijkstra74control,dolev2000book}.
Clock synchronisation tasks have been widely studied in both Byzantine-tolerant~\cite{halpern1984fault-journal,dolev1984possibility-journal,lundelius1984upper,lamport85clocks,srikanth1987optimal,welch1988new,bund2019fault,lenzen2022optimal} and self-stabilising~\cite{GoudaHerman1990Unison,AroraDolevGouda1991Clocks,couvreur1992asynchronous,AwerbuchKuttenMansourPattShamirVarghese1993Sync-journal,boczkowski2017minimizing,FeldmannKhazraeiScheideler2020Beeping,bastide2021self,dolev2021logarithmic,emek2021thin}~settings.

\subsection{The setting: clock synchronisation with extreme fault tolerance}

Disruptive events often cause both  transient and non-transient failures simultaneously, so a truly resilient algorithm has to cope with~both.
Remarkably, Dolev and Welch~\cite{dolev2004self} showed that it \emph{is} possible to obtain
clock synchronisation algorithms that recover from \emph{both} arbitrary transient faults (self-stabilisation) and tolerate $f < n/3$ arbitrary non-transient faults in message-passing systems with $n$ processes.
While the algorithms of Dolev and Welch used $2^{O(f)}$ expected time to stabilise, subsequent work has gradually improved the performance of both randomised and deterministic algorithms for
many self-stabilising synchronisation tasks.

In this line of research, the focus has been on two main problem variants: \emph{synchronous counting} (also known as digital clock synchronisation)~\cite{dolev2004self,hoch2006self,ben2008fast,dolev2016synchronous,lenzen2017efficient,lenzen2019near,lenzen2026early} and \emph{pulse synchronisation}~\cite{daliot2003self,daliot2003linear,dolev2007byzantine,dolev2014fault,lenzen2019self,khanchandani2019self}.
The problems are closely related, but typically studied in different timing models.
While in this work we consider only crash faults, as we aim to prove a strong lower bound, prior work on these problems has predominantly focused on upper bounds under the Byzantine fault setting.

\paragraph{Pulse synchronisation.}
The pulse synchronisation problem~\cite{daliot2003self,dolev2014fault}
is studied in the \emph{bounded-delay} model, where the message delivery times and local clock drift of processes are bounded by known constants, but adversarially controlled.
The goal is to synchronise the local clocks of processes so that all non-faulty processes produce an infinite sequence of synchronised pulses (i.e., \emph{logical} clock ticks) such that eventually after a finite stabilisation phase
\begin{enumerate}[label=(\roman*),noitemsep]
  \item  all non-faulty processes generate a pulse at roughly the same time, and
  \item the pulses are generated at a controlled frequency so that they are sufficiently far apart.
\end{enumerate}
The idea is that the pulses delimit logical synchronised rounds so that a synchronous system can be simulated.
See \Cref{fig:intro-fig-a} for an illustration.
With self-stabilisation, state-of-the-art deterministic algorithms can stabilise in $O(f)$ time and randomised algorithms in $\polylog f$ time~\cite{lenzen2019self}.

The challenge in pulse synchronisation comes from the combination of arbitrary initial states and non-transient faults.
If self-stabilisation is not required, i.e., the system is always cleanly initialised, then the problem is easy to solve in \emph{constant} time. Srikanth and Toueg~\cite{srikanth1987optimal} gave a simple algorithm with optimal skew and frequency while tolerating the optimal number $f < n/3$ of arbitrary non-transient faults.
If local clocks are initially closely synchronised, Welch and Lynch~\cite{welch1988new} showed that approximate synchronisation can be efficiently maintained by using approximate agreement over reals~\cite{dolev1986reaching} to correct local clocks fast.

\begin{figure}
  \centering
  \CountingPulseFigure[]
  \caption{Two self-stabilising synchronisation tasks under crash faults. The grey regions indicate the stabilisation phase during which outputs may be inconsistent. After this, outputs must satisfy the requirements of the task. In both examples, process 4 crashes during stabilisation.
    (a) In pulse synchronisation, the non-faulty processes output synchronised pulses in groups, each within a time window of length $\sigma$, and consecutive
  groups are sufficiently far apart. (b) In synchronous counting, processes must agree exactly on labels that are incremented by one modulo $q$ in each round. }\label{fig:intro-fig}{
    \phantomsubcaption\label{fig:intro-fig-a}
    \phantomsubcaption\label{fig:intro-fig-b}
  }
\end{figure}

\paragraph{Synchronous counting.}
In the exact synchronous counting problem~\cite{dolev2004self}, the assumption is that globally synchronous logical clock ticks are available (e.g., from an underlying pulse synchronisation algorithm), but these clock ticks need to be \emph{labelled} in a consistent manner to obtain round numbers.
More precisely, for a given integer $q \ge 2$, the constraints are that after stabilisation
\begin{enumerate}[label=(\roman*),noitemsep]
  \item exact agreement: the outputs of any two non-faulty processes are equal in each round, and
  \item progress: non-faulty processes increment their outputs by one modulo $q$ each round.
\end{enumerate}
\Cref{fig:intro-fig-b} illustrates the problem.
Synchronous counting has deterministic time complexity of $\Theta(f)$ rounds~\cite{hoch2006self,dolev2016synchronous}, and with randomisation
it can be solved in $O(1)$ expected rounds similarly to consensus~\cite{ben2008fast}.
Indeed, given that the problem requires \emph{exact} synchronisation, there exist reductions between consensus and synchronous counting both ways~\cite{hoch2006self,dolev2015fault,dolev2016synchronous}.
Subsequently, several papers have studied communication-efficient reductions to consensus in order to reduce the communication overhead of counting protocols~\cite{lenzen2017efficient,lenzen2019near,lenzen2026early}.

\subsection{The question: is approximate self-stabilising clock synchronisation hard?}

Unlike synchronous counting, whose time complexity matches that of consensus~\cite{fischer1982lower,aguilera1999simple}, no time lower bounds are known for pulse synchronisation.
Despite the long line of work improving the upper bounds for pulse synchronisation~\cite{daliot2003self,dolev2004self,dolev2007byzantine,dolev2014fault,dolev2015fault,lenzen2019self}, the progress has essentially been stuck at a seeming linear-in-$f$ deterministic stabilisation time~barrier.

The pulse synchronisation problem only asks for \emph{approximate} agreement on time, whereas synchronous counting requires \emph{exact} agreement.
Given that without self-stabilisation approximate clock synchronisation can be solved much faster~\cite{srikanth1987optimal,welch1988new}, this raises the question whether self-stabilising approximate clock synchronisation could be as easy as approximate agreement over reals~\cite{dolev1986reaching,fekete1990asymptotically,coan1988compiler,abraham2005optimal}.
In this work, we finally settle the following question~\cite{lenzen2019self}:
\begin{quotation}
  ``Is pulse synchronisation \emph{at least as hard} as synchronous consensus?''
\end{quotation}
The answer turns out to be  somewhat subtle. We show that pulse synchronisation and related approximate self-stabilising synchronisation tasks require $\Omega(f)$ deterministic time to stabilise even in the crash-fault model.
However, these self-stabilising tasks are \emph{slightly} easier than consensus: they have roughly the same time complexity as 2-set agreement and cycle agreement, both of which require $\lfloor f/2 \rfloor + O(1)$ rounds~\cite{chaudhuri2000tight,alistarh2023wait}. This is less than the $f+1$ rounds required for consensus.
Since Byzantine faults subsume crash faults, the lower bound for crash faults also applies to the previously studied Byzantine setting. In particular, this establishes the asymptotic time-optimality of existing deterministic Byzantine-tolerant algorithms for pulse synchronisation~\cite{daliot2003self,lenzen2019self}.

\section{Overview of results}\label{sec:overview}

We prove our lower bounds in a strong model of computing. We assume a fully connected and synchronous message-passing system, where processes can only fail by crashing. That is, the local clocks have no clock drift and the message delays are always bounded by a known constant.
Each process has a unique identifier  and these identifiers cannot be affected by transient faults.

All processes start with an arbitrary initial state, taken from a finite set of allowed states defined by the protocol.
In every communication round, each process (1) sends its current state to all processes, (2)
receives the states from all non-crashed processes, and (3) updates its local state and outputs a value.
Processes that crash in a round fail to send their messages to an adversarially chosen subset of processes. In all subsequent rounds, crashed processes never send any messages.
A process is \emph{non-faulty} in an execution if it never crashes in the execution.

\subsection{Lower bound for pulse synchronisation}

The pulse synchronisation problem has been previously investigated in the bounded-delay model with Byzantine faults~\cite{daliot2003linear,dolev2014fault,lenzen2019self}. As we prove a lower bound, we consider a simpler synchronous variant with crash faults that is at least as easy as the general non-synchronous version.
This is because executions in the bounded-delay model can also behave completely synchronously, and Byzantine faults subsume crash faults.

The output of a pulse synchronisation protocol is a binary indicator for each round $t$ that denotes for each process whether the process generates a pulse in round $t$ or not.
For integers $a \le b$, we write $[a,b] = \{a, \ldots, b\}$ for a closed interval and $(a,b) = \{a+1, \ldots, b-1\}$ for an open interval.

\begin{definition}[Pulse synchronisation]\label{def:pulse-sync}
  Let $\sigma \ge 1$ and $T \in \N$.
  A protocol is an \emph{$f$-resilient pulse synchronisation protocol with skew $\sigma$ that stabilises in $T$ rounds} if every execution with
  at most $f$ crashes has an infinite sequence of intervals $[a_0,b_0], [a_1,b_1], \ldots$ with $a_0 \le T$ such that for all $k \in \N$
  \begin{enumerate}[label=(P\arabic*),noitemsep]
    \item bounded skew: $b_k - a_k \le \sigma$, and each non-faulty process pulses exactly once in $[a_k, b_k]$, and
      \label{def:pulse-sync:skew}
    \item \label{def:pulse-sync:separation}
      pulse separation: $a_{k+1} > b_k + \sigma$, and no non-faulty process pulses in  $(b_k, a_{k+1})$.
  \end{enumerate}
\end{definition}
\Cref{fig:intro-fig-a} illustrates the problem for $\sigma=1$ in an execution that stabilises in four rounds.
The above problem relaxes the standard pulse synchronisation problem, see, e.g.,~\cite{lenzen2019self}. In our relaxation, the pulse separation condition~\labelcref{def:pulse-sync:separation} only asks for a \emph{lower bound} on when a new pulse can be generated; we do not require an upper bound for the time to generate a new pulse, apart from the first pulse, which must happen by round $T$. Our main result is a  lower bound of $\Omega(f)$ for the stabilisation time of any pulse synchronisation protocol:

\begin{restatable}{theorem}{pulsesynclb}\label{thm:pulse-synchronisation-lower-bound}
  Let $\sigma>0$ and $n>f+2$.
  There is no $f$-resilient finite-state  pulse synchronisation protocol with skew $\sigma$ that stabilises in $\lfloor f/2 \rfloor - 2\sigma$ rounds.
\end{restatable}

For $\sigma = 1$, the lower bound is tight up to an additive $O(1)$ term in the crash-fault setting, as shown later in \Cref{sec:upper-bound}.

\subsection{Lower bounds for approximate counting tasks}

We prove the lower bound for pulse synchronisation by a reduction from a certain \emph{approximate} relaxation of synchronous counting.
As this relaxation is somewhat technical, we build our intuition with perhaps the most obvious relaxation of synchronous counting. While this will not be precisely the problem we use for the reduction, it \emph{almost} captures everything we need.

\begin{figure}
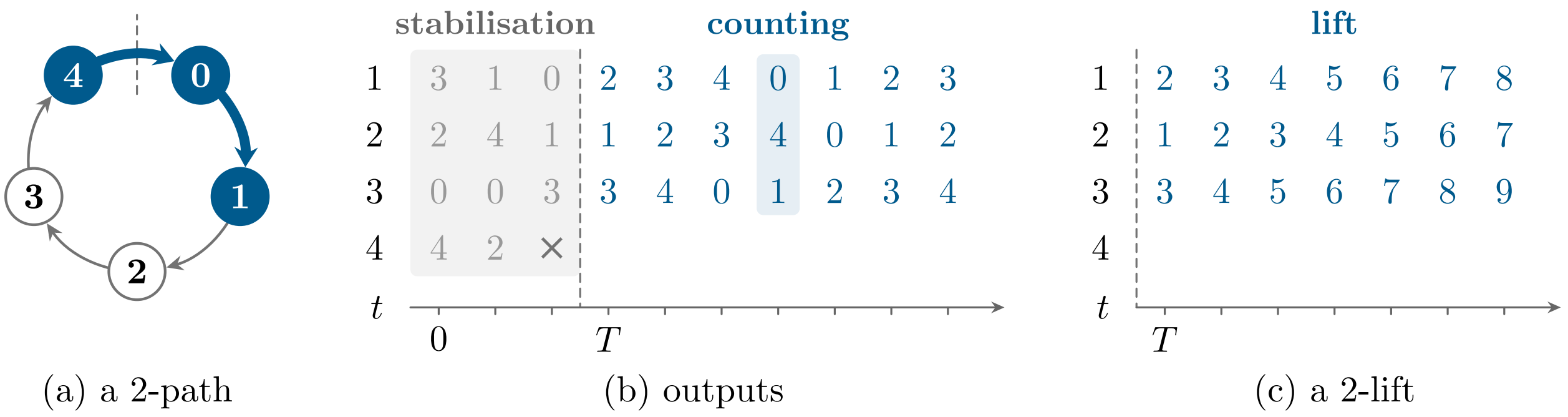

  \centering
  \PathCountingLiftFigure
  \caption{Approximate counting for $q=5$ and $\sigma=2$. (a) The cycle $\Z_5$ with the 2-path $\{4,0,1\}$. (b) An execution of a protocol that stabilises in round $T=3$. Process 4 crashes during stabilisation. After stabilisation, the outputs lie on a $\sigma$-path on $\Z_q$. The outputs in the blue column correspond to the $\sigma$-path given in the first panel. (c) A 2-lift of the outputs from $\Z_5$ to~$\Z$.}
  \label{fig:apx-counting}{
    \phantomsubcaption\label{fig:apx-counting-a}
    \phantomsubcaption\label{fig:apx-counting-b}
    \phantomsubcaption\label{fig:apx-counting-c}
  }
\end{figure}

\subsubsection{The first attempt: Approximate counting}
For $q \in \mathbb{N}$, we write $\Z_q = \{0, 1, \ldots, q-1\}$. For $\sigma \in \mathbb{N}$, we say that $I \subseteq \Z_q$ is a \emph{$\sigma$-path} if $I = \{c + k \bmod q \colon  0 \le k \le \sigma\}$ for some $c \in \Z_q$. If we interpret $\Z_q$ as the $q$-cycle graph, where there is an edge $(u,v)$ whenever $v = u +1 \bmod q$, then a $\sigma$-path is simply a path of length $\sigma$ on this cycle; see \Cref{fig:apx-counting-a}. We define $\sigma$-approximate $q$-counting as follows.

\begin{definition}\label{def:eager-counting}
  Let $\sigma \in \N$ and $q > \max\{2\sigma,1\}$.
  Consider a protocol with outputs $\phi(i,t) \in \Z_q$ and let $\Phi(t)$ be the set of outputs of non-faulty processes in round $t$.
  We say that the protocol is an $f$-resilient \emph{$\sigma$-approximate $q$-counting protocol} that  stabilises in $T$ rounds if every execution with at most $f$ crashes satisfies for all $t \ge T$ the following properties:
  \begin{enumerate}[label=(A\arabic*),noitemsep]
    \item $\sigma$-agreement: $\Phi(t)$ is contained in some $\sigma$-path of $\Z_q$, and \label{def:eager-counting:A1}
    \item progress: $\phi(i, t+1) = \phi(i,t) + 1 \bmod q$ for every non-faulty process $i$. \label{def:eager-counting:A2}
  \end{enumerate}
\end{definition}

\Cref{fig:apx-counting-b} illustrates the problem.
As in pulse synchronisation,
the parameter $\sigma$ controls the \emph{skew} of the outputs.
For $\sigma=0$ and any $q \ge 2$, we get the exact counting problem. For $\sigma \ge 1$ and $q>2\sigma$, the approximate counting problem can be seen as a ``self-stabilising'' version of the cycle agreement problem, that is, approximate agreement on a cycle graph, which requires $\lfloor f/2 \rfloor +1$ rounds~\cite{alcantara2019topology,alistarh2023wait,rybicki2026solvability}.
The condition $q > 2\sigma$ ensures that the shortest path between two outputs at distance at most $\sigma$ is unique, and we can determine which processes are ``ahead'' and which are ``behind''.

Given that synchronous counting lower bounds are obtained by reductions from consensus, one might expect that a similar reduction from cycle agreement to approximate counting would immediately give us a matching time lower bound.
Unfortunately, as explained later in \Cref{ssec:challenges}, such simple reductions fail due to issues related to self-stabilisation.
Another problem is that the separation property \labelcref{def:pulse-sync:separation} of pulse synchronisation allows the pulse gaps to ``vary'' by sometimes being longer and sometimes shorter. Hence, the progress condition of approximate counting is too strict for a naive reduction to work out. We remedy this by considering a further relaxation of the counting problem that still remains hard.

\subsubsection{The hard problem: Non-trivial counting}

We now introduce a new problem called \emph{non-trivial counting}. This is a fairly abstract-looking problem, but it captures the hardness of most natural variants of self-stabilising approximate clock synchronisation problems, including $\sigma$-approximate counting and pulse synchronisation.
We start by defining the notion of a \emph{$\sigma$-lift} that ``unwraps'' the outputs of non-faulty processes from $\Z_q$ to $\Z$ while preserving in $\Z$ the relative distances between the original outputs.

\begin{figure}
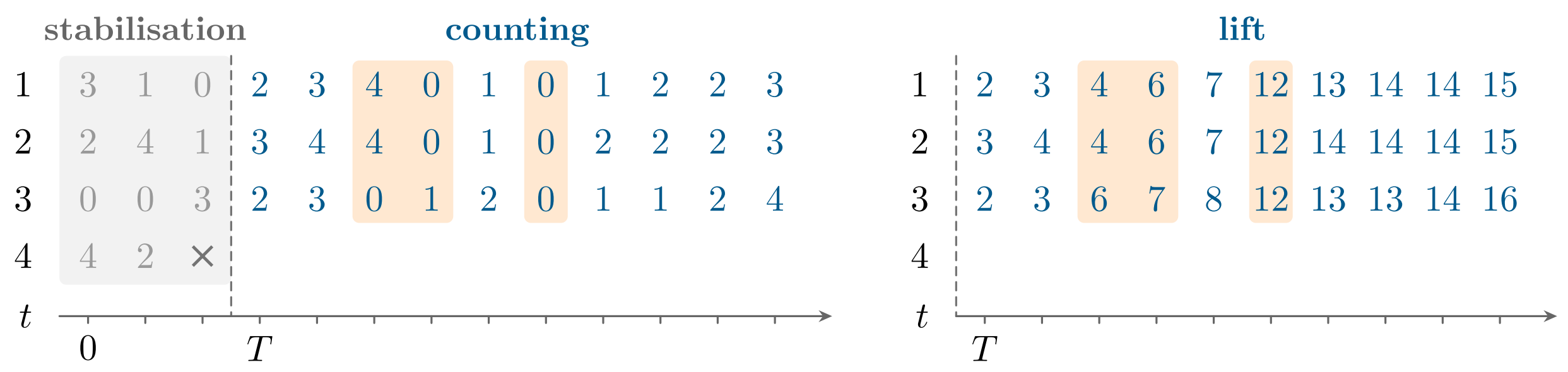

  \centering
  \CountingLiftFigure
  \caption{Non-trivial counting with $q=6$ and $\sigma=2$. (a) An execution that stabilises in round $T=3$. Process 4 is faulty. In the orange rounds, the outputs switch to 0. (b) A $\sigma$-lift of the outputs from round $T$~onwards. When the outputs reset to 0, the lift jumps to the next multiple of $q$.}
  \label{fig:lift}{
    \phantomsubcaption\label{fig:lift-a}
    \phantomsubcaption\label{fig:lift-b}
  }
\end{figure}

\begin{definition}\label{def:sigma-lift}
  Let $\sigma \in \N$ and $q > \max\{2\sigma,1\}$.
  Consider  an execution  with outputs $\phi(i,t) \in \Z_q$ and the set $P$ of non-faulty processes.
  We say that the map $\lift{\phi} \colon P \times \Z \to \Z$ is a \emph{$\sigma$-lift} of the outputs from round $T$ onwards if all rounds $t \ge T$ and non-faulty processes $i$ and $j$ satisfy
  \begin{enumerate}[label=(L\arabic*),noitemsep]
    \item compatibility: $\phi(i,t) = \lift{\phi}(i,t) \bmod q$, \label{def:sigma-lift:l1}
    \item $\sigma$-agreement: $|\lift{\phi}(i,t) - \lift{\phi}(j,t)| \le \sigma$, and \label{def:sigma-lift:l2}
    \item bounded increments: $0 \le \lift{\phi}(i,t+1) - \lift{\phi}(i,t) < q$. \label{def:sigma-lift:l3}
  \end{enumerate}
  That is, the lifted outputs agree with the original outputs modulo $q$, stay within distance $\sigma$ of each other, and never decrease, increasing by less than $q$ per round.
\end{definition}

\Cref{fig:apx-counting-c} illustrates a $\sigma$-lift for the approximate counting example. Compared to approximate counting, non-trivial counting is a much more relaxed problem:
after stabilisation, the outputs of non-faulty processes are allowed to, e.g., reset to 0, jump ahead by more than one, or stall arbitrarily long, as long as they have a $\sigma$-lift. Formally, the problem is defined as follows.

\begin{definition}\label{def:non-trivial-counting}
  Let $\sigma \in \N$ and $q > \max\{2\sigma,1\}$.
  We say that a protocol with outputs $\phi(\cdot) \in \Z_q$ is an $f$-resilient \emph{non-trivial $\sigma$-approximate $q$-counting protocol} that stabilises in $T$ rounds if
  \begin{enumerate}[label=(N\arabic*),noitemsep]
    \item \emph{every} execution with at most $f$ crashes has a $\sigma$-lift of the outputs from round $T$, and \label{def:non-trivial-counting:n1}
    \item in \emph{some} crash-free execution, the output of \emph{some} process changes infinitely often. \label{def:non-trivial-counting:n2}
  \end{enumerate}
\end{definition}
\Cref{fig:lift-a} illustrates outputs allowed by non-trivial counting and \Cref{fig:lift-b} gives a corresponding $\sigma$-lift.
We will again call $\sigma$ the skew of the protocol.
Observe that the above problem requires only very weak guarantees for the outputs. Condition \labelcref{def:non-trivial-counting:n1} asks that the (lifted) outputs remain at distance at most $\sigma$ and the outputs can increment each round by at most $q-1$ in executions with at most $f$ crashes. In particular, protocols can have outputs that in some rounds stall for arbitrarily long or ``jump'' by incrementing their outputs by more than one.
To ensure that such protocols have non-trivial outputs,  Condition \labelcref{def:non-trivial-counting:n2} only asks that in \emph{some} crash-free execution we can find a process that keeps making progress by changing its output.
We show the following result.

\begin{restatable}{theorem}{nontriviallb}\label{thm:non-trivial-lower-bound}
  Let $\sigma \in \N$, $q >\max\{2 \sigma,1\}$, $n>f+2$ and $f>0$. There is no $f$-resilient finite-state protocol for non-trivial $\sigma$-approximate $q$-counting that stabilises in $\lfloor f/2 \rfloor$ rounds.
\end{restatable}

As non-trivial counting reduces to both pulse synchronisation and approximate counting, we obtain an $\Omega(f)$ time lower bound for these (and other) tasks.

\subsection{Reductions from non-trivial counting}

We now show the reductions for both approximate counting and pulse synchronisation. We start with  approximate counting, where the reduction is almost immediate.

\paragraph{Non-trivial counting reduces to approximate counting.}
Suppose we have an approximate counting protocol that stabilises in $T$ rounds.
By the $\sigma$-agreement property \labelcref{def:eager-counting:A1}, we have that the outputs of non-faulty processes lie on some $\sigma$-path $I = \{c + k \bmod q \colon  0 \le k \le \sigma\}$ of $\Z_q$. For each non-faulty process $i$ with output $\phi(i,T) = c + k_i \bmod q$, where $k_i \in \{0, \ldots, \sigma\}$, we define the lift as
\begin{equation}\label{eq:canonical-lift}
  \lift{\phi}(i,t+1) :=
  \begin{cases}
    c + k_i & \text{if } t+1 \le T \\
    \lift{\phi}(i,t) +  \big(\phi(i,t+1)-\phi(i,t) \bmod q\big) & \text{otherwise}.
  \end{cases}
\end{equation}
The lift satisfies  \labelcref{def:sigma-lift:l1,def:sigma-lift:l3} by construction.

To verify that $\lift{\phi}$ gives a $\sigma$-lift of the outputs starting from round $T$ on, it only remains to check that \labelcref{def:sigma-lift:l2} is satisfied.
By the $\sigma$-agreement property \labelcref{def:eager-counting:A1}, the outputs of any two non-faulty processes $i$ and $j$, and hence also the lifts, start at most distance $\sigma$ apart in round $T$.
By the progress property \labelcref{def:eager-counting:A2} of approximate counting, we get that $\phi(i,t+1)-\phi(i,t) \bmod q = 1$, and both processes increment their outputs by one each round and the respective lifts keep their difference.
Thus, \labelcref{def:non-trivial-counting:n1} of non-trivial counting is satisfied.

Moreover, by the progress property \labelcref{def:eager-counting:A2} of approximate counting, it follows that in any crash-free execution the outputs increment by one each round; hence, property \labelcref{def:non-trivial-counting:n2} is satisfied. As the approximate counting protocol solves non-trivial counting, \Cref{thm:non-trivial-lower-bound} implies the following.

\begin{corollary}\label{corollary:apx}
  Let $\sigma \in \N$, $q>\max\{2\sigma,1\}$, $n > f+2$ and $f>0$. Then there is no $f$-resilient finite-state $\sigma$-approximate $q$-counting protocol that stabilises in $\lfloor f/2 \rfloor$ rounds.
\end{corollary}

\paragraph{Non-trivial counting reduces to pulse synchronisation.}
Suppose we have a pulse synchronisation protocol with skew $\sigma$ that stabilises in $T$ rounds. Recall that the output of a pulse synchronisation protocol indicates whether a pulse was generated in the given round or not.
We construct a non-trivial counting protocol with $q=2\sigma+1$.
The output $\phi(i,0)$ in the initial configuration is $0$ if the pulse synchronisation protocol starts in a state where process $i$ generates a pulse.
Otherwise, $\phi(i,0)$ is arbitrary.
In each round $t +1 >0$, every process $i$ performs the following:
\begin{itemize}[noitemsep]
  \item If process $i$ generates a pulse in round $t+1$,  then set $\phi(i,t+1) := 0$.
  \item Otherwise, set $\phi(i,t+1) := \min \{ 2\sigma, \phi(i,t) + 1\}$.
\end{itemize}
The above protocol solves a variant of approximate counting, where the non-faulty processes may in each round (a) increment their outputs by exactly one modulo $q$, (b) stall at the output $q-1 = 2\sigma$ for an arbitrarily long time (waiting for the next pulse) or (c) reset their outputs to 0.
It is easy to verify that because non-faulty processes generate a pulse infinitely often, the counting protocol satisfies condition \labelcref{def:non-trivial-counting:n2}: if a pulse is generated, the counter is reset to 0, and by \labelcref{def:pulse-sync:separation}, it is 1 in the next round.
It remains to show the property \labelcref{def:non-trivial-counting:n1}.
This step is a slightly tedious case analysis.

\begin{lemma}\label{lemma:lift}
  Any execution of the counting protocol with at most $f$ crashes has a $\sigma$-lift from round $T+2\sigma$ onwards.
\end{lemma}
\begin{proof}
  For a non-faulty process $i$, let  $\tau_i(k) \in [a_k,b_k]$ be the round when $i$ generates a pulse in the interval $[a_k,b_k]$, and define $\tau_i(-1) := T$. The properties
  \labelcref{def:pulse-sync:skew,def:pulse-sync:separation} of pulse synchronisation give
  \begin{enumerate}[noitemsep,label=(\roman*)]
    \item $|\tau_i(k) - \tau_j(k)| \le b_k-a_k \le \sigma$, and  \label{pulse-skew}
    \item $\tau_i(k+1) - \tau_j(k) \ge a_{k+1} - b_k > \sigma$ \label{pulse-separation}
  \end{enumerate}
  for all $k \ge 0$ and non-faulty processes $i$ and $j$.
  By \labelcref{def:pulse-sync:skew,def:pulse-sync:separation}, the rounds $\tau_i(0) < \tau_i(1) < \cdots$ are exactly the rounds $t \ge a_0$ in which
  process $i$ generates a pulse.
  For $t \ge a_0$, let $N_i(t)$ be the number of pulses generated by a non-faulty process $i$ in the interval $[a_0,t]$.
  Since the counters are reset when a pulse is generated, and the counters  increment only up to $2\sigma$, we have for all $t \ge T+2\sigma$ that
  \begin{equation}\label{eq:phi-k}
    \phi(i,t) = \min\{t - \tau_i(N_i(t)-1), 2\sigma \}.
  \end{equation}
  The above is $2\sigma$ whenever $N_i(t)=0$: in this case, process $i$ does not generate a pulse in the interval $[a_0,t]$, and $t-a_0 \ge t-T \ge 2 \sigma$. Hence, the counter of $i$ is not reset for at least $2\sigma$ rounds, and the right-hand side of \Cref{eq:phi-k} is $\min \{t-T,  2\sigma\} = 2\sigma$.

  For all $t \ge T + 2\sigma$, define the lift as $\lift{\phi}(i,t) := q N_i(t) + \phi(i,t)$.
  Since $\lift{\phi}(i,t) \bmod q = \phi(i,t)$, \labelcref{def:sigma-lift:l1} is satisfied.
  If there is a pulse in round $t+1$, the lift $\lift{\phi}(i,\cdot)$ increases by $q-\phi(i,t) < q$, as non-faulty $i$ cannot pulse in two consecutive rounds by \labelcref{pulse-separation}. Otherwise, it increases by at most one.
  Thus, \labelcref{def:sigma-lift:l3}~holds.
  To check \labelcref{def:sigma-lift:l2}, consider two non-faulty processes $i$ and $j$ in round $t \ge T+2\sigma$. There are three cases to consider:
  \begin{enumerate}
    \item Suppose $N_i(t)=N_j(t) = k$.
      Note that $|\min\{a,c\} - \min\{b,c\}| \le |a-b|$ for any $a,b,c \in \Z$.
      Using this and the definition of $\lift{\phi}$ yields
      \begin{align*}
        |\lift{\phi}(i,t) - \lift{\phi}(j,t)| &= |\phi(i,t) - \phi(j,t)| \\
        &= |\min\{t - \tau_i(k-1), 2\sigma \} -  \min\{t - \tau_j(k-1), 2\sigma \}|  \\
        &\le |t - \tau_i(k-1) - t + \tau_j(k-1)| \le \sigma,
      \end{align*}
      where the last inequality is trivial for $k=0$ and otherwise follows from \labelcref{pulse-skew}. Thus, \labelcref{def:sigma-lift:l2} holds.
    \item Suppose $N_i(t) = N_j(t)+1$ and let $k=N_j(t)$.
      Note that $t < \tau_j(k)$, so by \labelcref{pulse-skew}
      \begin{align*}
        m := t - \tau_i(k) < \tau_j(k) - \tau_i(k) \le \sigma.
      \end{align*}
      Hence $\phi(i,t) = m$ by \Cref{eq:phi-k}.
      If $k>0$, then \labelcref{pulse-separation} gives
      \begin{align*}
        t -  \tau_j(k-1) = m - \tau_j(k-1) + \tau_i(k) > m + \sigma.
      \end{align*}
      Otherwise, if $k=0$, then $t-\tau_j(-1)=t-T \ge 2\sigma > m +\sigma$.
      Since $m + \sigma + 1 \le 2\sigma$, we get $\phi(j,t) > m + \sigma$ by \Cref{eq:phi-k}.
      Using the definition of $\lift{\phi}$ yields
      \begin{align*}
        \lift{\phi}(i,t) - \lift{\phi}(j,t) &= q + \phi(i,t) - \phi(j,t)
        < q + m - (m + \sigma) = q - \sigma = \sigma + 1.
      \end{align*}
      Since $\phi(i,t) \ge 0$ and $\phi(j,t) \le 2\sigma < q$, we also have $\lift{\phi}(i,t) > \lift{\phi}(j,t)$, so
      \labelcref{def:sigma-lift:l2} follows.
    \item Suppose $N_i(t)-N_j(t) \ge 2$ and let $k := N_j(t)$. Then $\tau_i(k+1) \le t$.
      By \labelcref{pulse-separation}, we have $\tau_j(k) < \tau_i(k+1)$.
      Thus, $N_j(t) \ge k+1$, a contradiction, so this case cannot happen.
  \end{enumerate}
  The other cases with roles of $i$ and $j$ swapped are symmetric.
\end{proof}

With the above technical lemma, we are ready to show the pulse synchronisation lower bound.

\pulsesynclb*
\begin{proof}By assumption, the pulse synchronisation protocol stabilises in $T$ rounds.
  \Cref{lemma:lift} implies that there is a $\sigma$-lift from round $T+2\sigma$ onwards, so \labelcref{def:non-trivial-counting:n1} of non-trivial counting is satisfied.
  By \Cref{thm:non-trivial-lower-bound}, we get that $T+2\sigma > \lfloor f/2 \rfloor$, which gives the claimed lower bound.
\end{proof}

\subsection{Main challenges}\label{ssec:challenges}

As we have seen above, natural approximate synchronisation tasks have reductions from non-trivial counting.
Before we proceed to the proof outline of \Cref{thm:non-trivial-lower-bound}, we discuss why the lower bounds for pulse synchronisation and non-trivial counting do not easily follow from existing standard arguments used for self-stabilising synchronous counting and for terminating approximate agreement tasks.

\paragraph{Challenge 1: Simple reductions fail.}
A promising strategy is to (1) reduce the approximate counting problem to pulse synchronisation, and then, (2) reduce cycle agreement to approximate counting in order to get the desired lower bound. This was essentially one motivation for studying round complexity lower bounds for cycle agreement~\cite{nowak2019byzantine,alistarh2023wait}.
As we saw, the first step works out by choosing an appropriate relaxation of approximate counting, but the second provides a more serious obstacle.
To see why, recall the standard reduction of consensus to exact counting~\cite{dolev2016synchronous}:
\begin{enumerate}
  \item
    Assume $A$ is an exact $q$-counting protocol that stabilises in $T$ rounds.
  \item Take \emph{stabilised} configurations $C^{(b)} = (C_1^{(b)}, \ldots, C_n^{(b)})$ for each $b \in  \{0,1\}$, where $C_i^{(b)}$ is the state of process $i$ and in the configuration $C^{(b)}$ all processes output $b$.
  \item  To solve binary consensus,  each process $i$ with input $b \in \{0,1\}$ initialises its state in $A$ to be $C_i^{(b)}$ and simulates $A$ for $T$ rounds. If $A$ outputs $y_i \in \Z_q$ for process $i$ at the end of round $T$, then process $i$ sets $z_i := y_i - T \bmod q$ and outputs $z_i \bmod 2$.
\end{enumerate}
It is easy to verify that this satisfies agreement and validity of binary consensus, and hence, we get that $T \ge f+1$ from standard consensus lower bounds, e.g.,~\cite{fischer1982lower,aguilera1999simple,chaudhuri2000tight}.

Why does the same approach fail when trying to reduce, e.g., cycle agreement to 1-approximate counting? We could now try to define stabilised configurations $C^{(0)}, \ldots, C^{(q-1)}$, where the outputs agree. Suppose now that the inputs lie on an edge $\{0,1\}$, so that agreement is satisfied (outputs are within distance one on a cycle) and we must maintain validity of cycle agreement (outputs are a subset of the inputs). If we do the same simulation argument, where some processes initialise their state by taking values from $C^{(0)}$ and others by taking their state from $C^{(1)}$, then this is no longer necessarily a stabilised configuration with outputs $\{0,1\}$. The reason is that arbitrarily ``gluing'' states from different stabilised configurations is not guaranteed to yield a stabilised configuration.

Hence, when started on an initial configuration constructed this way, the approximate counting protocol may stabilise in $T$ rounds onto \emph{any} possible edge, and we cannot map it back to the input edge. This violates validity of cycle agreement.
One can try to wiggle out of this issue by choosing a larger set of stabilised configurations, such as those with outputs in $\{0,1\}, \{1,2\}, \ldots$ from which to choose the inputs, but then it is unclear which configurations to use when initialising the simulation. Thus, the reduction is faced with essentially needing to solve 2-set agreement \emph{before} the simulation.

\paragraph{Challenge 2: Lack of topological tools for self-stabilisation.}
If one conjectures that approximate counting or pulse synchronisation behave similarly to $2$-set agreement or cycle agreement, then one might be persuaded to learn and extend the tools used to prove lower bounds for these tasks. As it turns out, essentially all such techniques boil down to using arguments from combinatorial topology one way or another~\cite{chaudhuri2000tight,herlihy1999topological,saks2000wait,herlihy2013distributed,alcantara2019topology,alistarh2023wait,herlihy2009axiomatic,alistarh2019why-journal}.

Unfortunately, unlike for classic \emph{terminating} agreement problems, such as consensus, set agreement and cycle agreement, there are no readily available tools for characterising solvability or complexity of \emph{self-stabilising} tasks for which protocols \emph{can never terminate}.
For terminating tasks, one can always obtain the protocol complex describing the entire execution after applying finitely many topological operations (e.g., barycentric or chromatic subdivisions) to the input complex. But what is even the right topological object to look at for self-stabilising tasks where the protocols have infinite executions?

The most relevant recent work uses techniques based on point-set topology to characterise solvability of \emph{stabilising} consensus tasks in infinite executions~\cite{schmid2026topological}. However, these do not seem to easily extend to the self-stabilising setting, where the initial states can be arbitrary. Furthermore, unlike for stabilising consensus, where the outputs have to eventually agree on a single common value, in clock synchronisation the outputs \emph{have to} change over time to guarantee progress.

\section{The high-level argument and main ideas}

Our general approach is to use the combinatorial topology framework of distributed computing~\cite{herlihy2013distributed}.
We now explain the main ingredients of the lower bound proof that allow us to extend topological arguments to the self-stabilising setting with crash faults.
We assume finite-state protocols, and without loss of generality, assume that the state encodes the identity of the process.

To keep things simple, we now outline the argument in the special case of approximate counting. The general lower bound for non-trivial counting follows essentially the same approach; the generalisation mostly involves more careful bookkeeping and some additional technical details that need to be handled when the clocks may stall or jump ahead rather than incrementing by one in each round.

\subsection{Background: topology of synchronous computing}

A reader who is not familiar with the language of combinatorial topology may find it useful to skip ahead and read the topological preliminaries given in \Cref{sec:preliminaries} before reading the outline.
An \emph{abstract simplicial complex} $X$ is a collection of finite subsets of a set $V(X)$ with the property that if $A \in X$, then $B \in X$ for all $\emptyset \neq B \subseteq A$. The elements of $V(X)$ are called the \emph{vertices} of $X$ and the elements of $X$ the \emph{simplices} of $X$.

Herlihy, Rajsbaum and Tuttle~\cite{herlihy2009axiomatic} defined the synchronous full-information protocol complex $R(r,k)$ for $r$-round executions with \emph{at most} $k$ crashes per round.
Each vertex of this complex is an $r$-round view of the full-information protocol
(i.e., the history of the messages received by a process). The simplices of this complex are (sub)configurations that are reachable in some $r$-round execution. Formally, this complex is a union of combinatorial pseudospheres; see \Cref{ssec:topology}.

While Herlihy, Rajsbaum and Tuttle~\cite{herlihy2009axiomatic} considered terminating protocols, the synchronous full-information protocol complex can be adapted to the self-stabilising setting. The main issue  with self-stabilising protocols is that we do not have inputs per se. However, we treat the initial state of the process as its input for the full-information protocol. Hence, the allowed inputs are simply all possible initial configurations.
Herlihy, Rajsbaum and Tuttle bounded the homotopical connectivity of the complex $R(r,k)$. The main topological result we need follows from their work.

\begin{theorem}[Herlihy, Rajsbaum and Tuttle~\cite{herlihy2009axiomatic}]
  \label{thm:hrt}
  If $n > (r+1)k$, then the synchronous full-information protocol complex $R(r,k)$ is homotopically $(k-1)$-connected.
\end{theorem}

We only need the case when the complexes are \emph{path connected} (0-connected) and \emph{simply connected} (1-connected). Path connectivity means that the underlying graph $(V(X),E(X))$ of $X$ given by the edge set $E(X) = \{ e \in X : |e| = 2\}$ is connected in the graph-theoretic sense. Simple connectivity means that the complex is path connected and every edge loop (i.e., a closed walk in the underlying graph) can be contracted to a single vertex. The precise definitions appear in \Cref{sec:preliminaries}.

\subsection{Warmup: lower bound for exact non-trivial counting}

As a gentle warmup, let us sketch a time lower bound for \emph{exact} non-trivial counting, that is, the variant with skew $\sigma=0$.
This relaxes synchronous counting by allowing outputs to stall or increment possibly by more than one but less than $q$ steps. However, the outputs must always agree exactly.

\begin{theorem}\label{thm:warmup}
  Let $q \ge 2$, $f>0$ and $n > f+1$.
  There is no $f$-resilient protocol for exact non-trivial counting that stabilises in $f$ rounds.
\end{theorem}
\begin{proof}
  Suppose there exists a non-trivial counting protocol with skew $\sigma=0$ that stabilises in $T := f$ rounds.
  Let $X := R(T,1)$ be the synchronous full-information protocol complex whose vertices consist of $T$-round views in executions where at most one process crashes per round.
  By assumption the protocol stabilises in $T$ rounds. Hence, each simplex corresponds to a stabilised (sub)configuration that may occur in an execution of the protocol.
  We write $\out(v) \in \Z_q$ for the output of any process $i$ with view $v \in V(X)$.
  The 0-agreement property \labelcref{def:sigma-lift:l2} implies that
  \begin{equation}\label{eq:warmup-equality}
    \out(u) = \out(v)
  \end{equation}
  for any $\{u,v\}\in X$. In particular, this holds for any edge $\{u,v\} \in E(X)$.
  By \Cref{thm:hrt}, the synchronous full-information protocol complex $X$ is path connected, because
  \(
    n > T + 1 = f + 1
  \).
  Hence, we get that $\out(\cdot)$ is constant on $V(X)$, as the outputs agree along any path of $X$.

  Consider an infinite crash-free execution $(\rho_t)_{t \ge 0}$ and a process $i$ given by \labelcref{def:non-trivial-counting:n2}.
  Let $v$ be the $T$-round view of process $i$ in the crash-free execution started from $\rho_0$.
  By  \labelcref{def:non-trivial-counting:n2} there exists a round $t > T$ when process $i$ changes its output for the first time after round $T$. Let $v'$ be the $T$-round view of $i$ in the crash-free execution started from the configuration at time $t-T$.
  Since both $v$ and $v'$ are $T$-round views of the process $i$, they are vertices of $X$.
  Now
  \(
    \out(v) \neq \out(v')
  \), contradicting that $\out(\cdot)$ is constant on $V(X)$.
\end{proof}

\subsection{The lower bound idea for approximate counting}

Let us now consider how the lower bound works for $\sigma$-approximate $q$-counting, a special case of non-trivial counting.
me we have an $f$-resilient protocol for approximate counting that stabilises in $T := \lfloor f/2 \rfloor$ rounds.
Let $X := R(T,2)$ be the synchronous full-information protocol complex. By \Cref{thm:hrt}, the complex $X$ is simply connected,
as $n > f+2 \ge 2(T+1)$.
The vertices of $X$ consist of all the possible $T$-round views that can occur in executions with at most two crashes per round. Technically, the vertices of $X$ are not the \emph{states} in which processes are after $T$ rounds, but the states are a function of these views.
Since the state space and $T$ are finite, the complex $X$ is finite.

The simplices of $X$ correspond to (sub)configurations reachable in $T$-round executions in which at most 2 processes crash per round. Since $2T \le f$, the protocol stabilises in any such execution. Hence, every simplex represents a \emph{stabilised} (sub)configuration.
Since all configurations are stabilised, the outputs $\{\out(v) : v \in \rho \}$ for any simplex $\rho \in X$ lie on a $\sigma$-path of $\Z_q$. In particular, they are at most distance $\sigma$ apart on the $q$-cycle $\Z_q$ by the $\sigma$-agreement property \labelcref{def:eager-counting:A1} of approximate counting.

\paragraph{Step 1: Clock offsets give edge gradients.}
We define for every $\{u,v\} \in X$ the \emph{clock offset} $D(u,v)$ as
the \emph{signed} distance from $\out(u)$ to $\out(v)$ on the $q$-cycle $\Z_q$ so that the sign is positive if the shortest path goes clockwise (increasing values), and negative otherwise.
Since $\{u,v\} \in X$ is contained in a stabilised configuration, we get that
$D(u,v) \in \{-\sigma, \ldots, \sigma\}$ by the $\sigma$-agreement property~\labelcref{def:eager-counting:A1}.
This is well-defined: as $q > 2\sigma$, the shortest path between outputs is unique.
We also show that the map $D$ satisfies a \emph{triangle property}, i.e., that $D(u,v) + D(v,w) = D(u,w)$ for any $\{u,v,w\} \in X$.
This is because the outputs on any triangle lie on a $\sigma$-path, and summing their offsets adds up to 0.

\begin{figure}
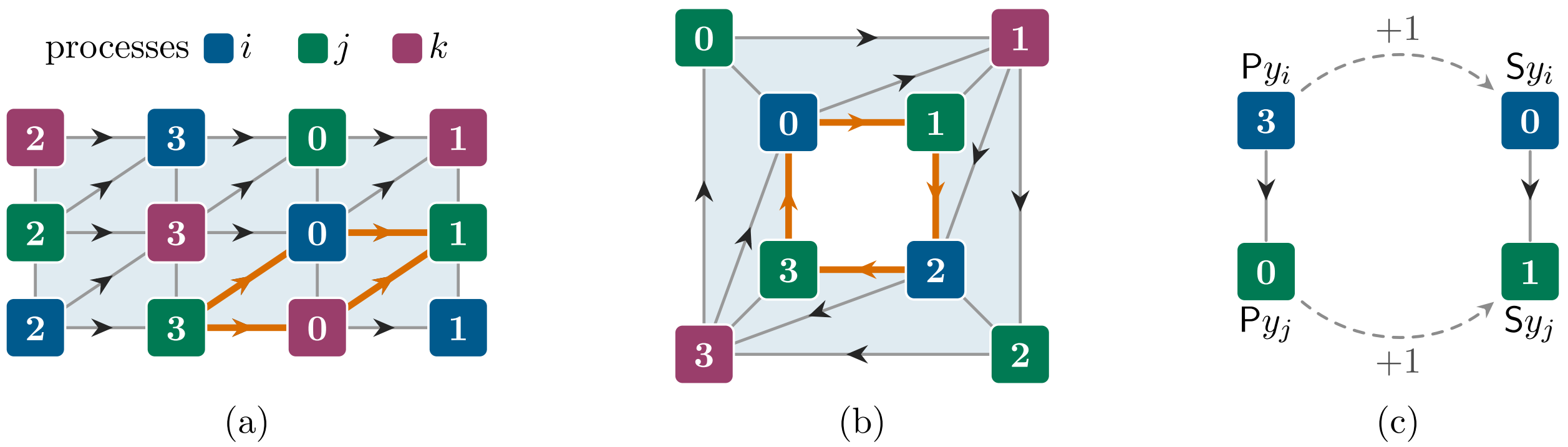

  \centering
  \GradientWindingComparisonFigure
  \caption{Clock offsets for $\sigma=1$ and $q=4$.
    Vertices are $T$-round views, coloured by process and labelled by output; filled triangles are
    configurations. Edges are oriented from $u$ to $v$ if $D(u,v)=1$ and unoriented if $D(u,v)=0$.
    (a) A simply connected protocol subcomplex, where the clock offsets give an edge gradient: offsets on any loop, such as the highlighted one, sum to 0.
    (b) A subcomplex that is not simply connected: the offsets on the highlighted loop sum to 4, so the clock offsets do not give an edge gradient.
    (c) The outputs of the prefix $\prefix{y_j}$ and suffix $\suffix{y_j}$ of a $(T+1)$-round view $y_j$ of process $j$ differ by one (dashed arrows): in approximate
    counting, both processes advance their outputs by exactly one modulo $q$. This gives $D(\prefix{y_i},\prefix{y_j}) = D(\suffix{y_i},\suffix{y_j})$ in the complex $X$.
  }
  \label{fig:gradient-illustration}{
    \phantomsubcaption\label{fig:gradient-illustration-a}
    \phantomsubcaption\label{fig:gradient-illustration-b}
    \phantomsubcaption\label{fig:gradient-illustration-c}
  }
\end{figure}

Recall that by \Cref{thm:hrt}, the complex $X$ is simply connected. Since the map $D$ satisfies the triangle property, this implies that $D$ is an \emph{edge gradient} of the complex $X$ (i.e., a simplicial 1-coboundary); see \Cref{lemma:from-cocycle-to-gradient}.
This simply means that there exists a map $\vartheta \colon V(X) \to \Z$ such that $D(u,v) = \vartheta(v) - \vartheta(u)$. Such a map $\vartheta$ is commonly called a \emph{potential}.
\Cref{fig:gradient-illustration-a} illustrates an edge~gradient and \Cref{fig:gradient-illustration-b} illustrates a complex where clock offsets do not give an edge gradient.

Since $X$ is also path connected, we show how to find a nice potential $\ell \colon V(X) \to \Z$ that lifts the outputs from $\Z_q$ to $\Z$ in the sense that the map satisfies
\begin{equation}\label{eq:D-potential}
  D(u,v) = \ell(v) - \ell(u) \qquad \text{ and } \qquad \ell(v) \equiv \out(v) \pmod q
\end{equation}
for all $\{u,v\} \in X$.
The key step is to constrain how these clock offsets behave using another complex.

\paragraph{Step 2: The clock offsets have to be globally consistent.}
For approximate counting,  the progress condition \labelcref{def:eager-counting:A2} forces the outputs of the processes to behave predictably: the output in round $T+1$ is the output in round $T$ incremented by one modulo $q$.

We now come to one of the key new ideas of the lower bound proof.
To get access to the \emph{temporal behaviour} of the self-stabilising protocol, we construct \emph{another} complex $Y = R(T+1,1)$, whose vertices are $(T+1)$-round views of executions in which at most one process crashes per round. Since $T + 1 \le f$, these executions have at most $f$ crashes, so the simplices of $Y$ are also stabilised subconfigurations.

We link $Y$ to $X$ as follows.
For any vertex $y \in V(Y)$, we define its \emph{prefix} $\prefix{y}$ as the $T$-round view without the $(T+1)$th round. Similarly, we define the \emph{suffix} $\suffix{y}$ of $y$ as the $T$-round view without the first round. The formal definitions are in \Cref{sec:preliminaries}.
Both the prefix $\prefix{y}$ and the suffix $\suffix{y}$ are $T$-round views, and hence, vertices of the complex $X$ we constructed above.
In fact, the prefix and suffix operations give chromatic simplicial maps from $Y$ to $X$.
This means that for any simplex $\tau \in Y$, the sets $\{\prefix{v} : v \in \tau\}$
and $\{\suffix{v} : v \in \tau\}$ are simplices of $X$.

The progress condition \labelcref{def:eager-counting:A2} of approximate counting ensures that the clock offsets of two processes $i$ and $j$ have to be globally consistent in the sense that
\[
  D(\prefix{y_i}, \prefix{y_j}) = D(\suffix{y_i}, \suffix{y_j})
\]
for any $\{y_i, y_j\} \in Y$; see \Cref{fig:gradient-illustration-c}.
Writing
\(
  \epsilon(y) := \ell(\suffix{y}) - \ell(\prefix{y}) - 1
\)
and using \Cref{eq:D-potential} yields
\begin{align*}
  \epsilon(y_i) - \epsilon(y_j) =  D(\prefix{y_i}, \prefix{y_j}) - D(\suffix{y_i}, \suffix{y_j})  = 0.
\end{align*}
In particular, $\epsilon$ is a constant map on $V(Y)$, since the complex $Y$ is path connected by \Cref{thm:hrt}. By the second condition of \Cref{eq:D-potential} and the fact that $\out(\suffix{y})=\out(\prefix{y})+1\bmod q$, we get that $\epsilon \equiv 0 \pmod q$.
Consequently, there exists some constant $c \in \Z$ such that the lift $\ell$ satisfies
\begin{equation}\label{eq:ell-suffix}
  \ell(\suffix{y}) - \ell(\prefix{y}) = 1 + \epsilon(y) = 1 + qc.
\end{equation}

\paragraph{Step 3: Global consistency implies an infinite complex.}
Finally, we look at what a fixed process $i$ does in an infinite crash-free execution.
For all $t \ge 0$, let $x_t$ be the $T$-round view of process $i$ when starting the execution from the configuration at time $t$.
Since the vertices of the complex $X$ are all possible $T$-round views of the protocol in executions with \emph{at most} two crashes per round, which include crash-free executions, we get that each $x_t \in V(X)$.
Now we have an infinite sequence $(x_t)_{t \ge 0}$ of $T$-round views. Of course, since $X$ is finite, some vertices are repeated in $(x_t)_{t \ge 0}$.

Since each $x_{t+1}$ and $x_t$ are the suffix and prefix of some $(T+1)$-round view, we get from \Cref{eq:ell-suffix} and some algebraic manipulation that
\(
  \ell(x_t) = \ell(x_0) + \left(1 + qc \right)t.
\)
No matter the value of $c \in \Z$, $\ell(x_t)$ diverges as $t \to \infty$, because $q \ge 2$.
Thus, the map $\ell \colon V(X) \to \Z$ takes on infinitely many different values, which contradicts the finiteness of the complex $X$.

This is essentially the whole argument for the lower bound construction.
Extending this argument to cover non-trivial counting requires mainly a more careful handling of the clock offsets and their increments in the second step. The full details appear in \Cref{sec:non-trivial-lower-bound}.

\subsection{Tightness of the lower bounds}

Existing deterministic Byzantine-tolerant exact synchronous counting protocols solve approximate counting in asymptotically optimal $\Theta(f)$ rounds~\cite{hoch2006self,lenzen2019near}.
Moreover, it is not too hard to come up with a protocol that matches \Cref{corollary:apx} up to an $O(\log q)$ additive term, so the lower bound cannot be improved much in our synchronous setting with crash faults.

The general idea is as follows.
If any round has at most one crash, then the processes receive at most two different sets of messages in that round; by the pigeonhole principle,
this happens within the first $\lfloor f/2 \rfloor + 1$ rounds. Afterwards, there are at most three output values and they lie on a short path of the cycle, and running approximate agreement on a path halves the distance between outputs in every round. The details are in \Cref{sec:upper-bound}.

This construction shows that with $q=4$ and $\sigma=1$, we obtain an approximate counting protocol that exactly matches the lower bound of \Cref{corollary:apx} for any even $f$.
We can now solve pulse synchronisation by generating a pulse whenever the output is congruent to 0 modulo 4. This matches \Cref{thm:pulse-synchronisation-lower-bound} up to an additive $O(1)$ term.

\subsection{Outline of the paper}

The remainder of the paper is organised as follows.
\Cref{sec:further} discusses further related work.
\Cref{sec:preliminaries} introduces the topological tools used to prove the lower bound.
We favour simple ``combinatorial'' definitions and elementary arguments, rather than invoking geometric realisations of topological objects or cohomology theory. We hope this makes the proof easy to follow for readers with limited background in algebraic topology.
\Cref{sec:non-trivial-lower-bound} gives the general lower bound for non-trivial counting.
\Cref{sec:upper-bound} discusses upper bounds. \Cref{sec:conclusions} concludes the paper.

\section{Further related work}\label{sec:further}

Combinatorial and algebraic topology plays a key role in characterising the solvability and complexity of fault-tolerant \emph{terminating} distributed agreement tasks~\cite{herlihy2013distributed}.
Topological arguments are the main tool for tasks where exact agreement is not required, such as set agreement~\cite{chaudhuri2000tight,saks2000wait,herlihy1999topological,fraigniaud2025lower}, loop agreement~\cite{herlihy2003classification,liu2009classifying}, and approximate agreement on graphs and simplicial complexes~\cite{alcantara2019topology,alistarh2023wait,liu2023impossibility,rybicki2026solvability,albert2026solving}.

\paragraph{Synchronous topology.}
\citet{fraigniaud2025lower} extended the homotopical connectivity-based analysis
of synchronous $k$-set agreement
in~\cite[Chapter~13]{herlihy2013distributed} to arbitrary communication
networks, while correcting a gap in the carrier-map framework.
Their techniques may prove useful for extending the clock synchronisation bounds to general communication networks.
However, their construction uses executions with \emph{exactly} $k$ crashes per
round, whereas we use the complexes of \cite{herlihy2009axiomatic} that permit \emph{at most}
$k$ crashes per round. In our proof, we need to access crash-free executions as well.

\citet{schmid2026topological} investigated stabilising consensus using a point-set topology that captures infinite executions under message adversaries.
However, in stabilising consensus, the initial states are not arbitrary.
Moreover, stabilising consensus requires eventual agreement on a stable common value, whereas our tasks only need approximate agreement \emph{but} require that the outputs change infinitely often (in some execution).

\paragraph{Gradients and cohomological approaches.}
The general idea of clock offsets being gradients already appears in prior work~\cite{boulinier2008synchronous,hylton2023toward}.
For example, \citet{boulinier2008synchronous} studied self-stabilising clock synchronisation without non-transient faults. They essentially gave the same local offset identity as we use in building the gradient,
and similarly used the sums of gradients across paths.
However, they define gradients on the \emph{communication network}.
In our proof, the edge gradient is defined over the \emph{protocol complex} that captures all executions of the protocol.

Only a few explicit cohomological approaches have been proposed for distributed agreement tasks. Recently, \citet{felber2025brief} introduced a sheaf-theoretic approach to characterising solvability of terminating tasks.
\citet{hylton2023toward} identified an interesting connection between algebraic topology and clock synchronisation: similarly, they
interpret clock offsets using cochains (edge cocycles) and coboundaries (edge gradients). They also discuss a sheaf-theoretic approach to time
synchronisation. However, their work does not consider time complexity lower bounds.
Moreover, our work only requires the elementary notions of edge cocycles and edge coboundaries.

\citet{cai2014cyclic} and \citet{gravner2018limiting} studied synchronisation phenomena of generalised cellular automata on arbitrary graphs. Both analyses of the Greenberg--Hastings model used a similar edge gradient and path integration (i.e., summing the gradient over a walk) approach. While our lower bound argument uses a similar idea, it does so in the protocol complex. Moreover, unlike the analyses on networks, we ensure compatibility of clocks over time by using two protocol complexes that are related by chromatic simplicial maps.

\section{Topological preliminaries}\label{sec:preliminaries}

\subsection{Abstract simplicial complexes}

Let $X$ be a finite collection of finite sets.
We say that $X$ is an \emph{abstract simplicial complex} if
\(
  \rho \in X \implies \tau \in X \text{ for all non-empty } \tau \subseteq \rho
\).
That is, $X$ is closed under taking non-empty subsets.
An element $\rho \in X$ is called a \emph{simplex} of $X$.
The \emph{dimension} of a simplex $\rho \in X$ is
\(
  |\rho|-1
\).
Simplices of dimension zero correspond to \emph{vertices} of $X$.
We write $V(X) = \{ v : \{v\} \in X\}$ for the set of vertices of $X$.
A simplex of dimension one is  an \emph{edge}.
We write $E(X)$ for the set of edges of $X$.
A simplex of dimension two is  a \emph{triangle}.
The graph $(V(X), E(X))$ is the \emph{underlying graph}~of~$X$.

For a collection of sets $Y$, we say that \(
  \{  \tau : \emptyset \neq \tau \subseteq \rho, \rho \in Y \}
\)
is the simplicial complex generated by $Y$.
For two simplicial complexes $X$ and $Y$, we say that a map $F \colon V(Y) \to V(X)$ is \emph{simplicial} if for any simplex $\rho \in Y$ we have $F(\rho) \in X$, where $F(\rho) = \{  F(v) : v \in \rho \}$.

\subsection{Connectivity of abstract simplicial complexes}

The \emph{homotopical connectivity} of an abstract simplicial complex $X$ is defined in terms of the homotopy groups of the geometric realisation of $X$ as follows: an abstract simplicial complex $X$ is \emph{$k$-connected} if the first $k+1$ homotopy groups of the geometric realisation of $X$ are trivial~\cite{armstrong1983topology,hatcher2002algebraic}.
However, for our proof, it suffices to only work with lower orders of connectivity, namely path connectedness (0-connectivity) and simple connectedness (1-connectivity), which can be defined combinatorially without needing to introduce homotopy theory. We opt for this approach.

\paragraph{Edge paths and edge loops.}
Let $X$ be an abstract simplicial complex.
An \emph{edge path} in $X$ is a sequence $\alpha = (v_0, \ldots, v_k)$ of (possibly repeated) vertices such that $\{v_i, v_{i+1}\} \in X$ for all $0 \le i <k$.
If $v_0 = v_k$, then we say that $\alpha$ is an \emph{edge loop}.
A \emph{contraction} of the edge path $\alpha$ is an edge path $\beta$ obtained by removing vertices using the following rules:
\begin{enumerate}[label=(R\arabic*),noitemsep]
  \item \label{rule:remove-identical} If $v_i = v_{i+1}$, then $\beta = (v_0, \ldots, v_i, v_{i+2}, \ldots, v_k)$ is a contraction of $\alpha$.
  \item \label{rule:triangle-contraction} If $\{v_i,v_{i+1}, v_{i+2}\}\in X$, then  $\beta =(v_0, \ldots, v_i, v_{i+2}, \ldots, v_k)$ is a contraction of $\alpha$.
\end{enumerate}
Two edge paths $\alpha$ and $\beta$ are \emph{equivalent} if we can get from one to the other using a finite sequence of contractions or their inverses.
An edge loop starting from $v_0$ is \emph{contractible} (that is, topologically, null-homotopic) if it is equivalent to the edge path consisting only of the vertex $v_0$.

\paragraph{Connectivity.}
A complex $X$ is \emph{path connected} (0-connected) if for any two vertices $u,v \in V(X)$ there is an edge path between $u$ and $v$.
That is, the underlying graph of the complex is connected in the graph-theoretic sense.
A path connected simplicial complex $X$ is \emph{simply connected} (1-connected) if every edge loop of $X$ is contractible.
This is equivalent to the condition that the geometric realisation of $X$ has a trivial fundamental group~\cite[Theorem 6.10]{armstrong1983topology}.

\paragraph{Edge gradients.}
An \emph{edge cocycle} of $X$ is a map $\omega$ that assigns an integer $\omega(u,v)$ to every ordered pair $(u,v)$ of vertices, where $\{u,v\} \in
X$, so that the following \emph{triangle property}
\begin{equation}
  \omega(u,v) + \omega(v,w) = \omega(u,w) \tag{$\blacktriangle$} \label{eq:triangle-property}
\end{equation}
is satisfied for all $\{u,v,w\} \in X$. This implies that $\omega(u,u)=0$ and that $\omega$ is antisymmetric, i.e., $\omega(u,v)=-\omega(v,u)$. We say that $\omega$ is an \emph{edge gradient} if there exists some $\vartheta \colon V(X) \to \Z$ such that
\[
  \omega(u,v) = \vartheta(v) - \vartheta(u)
\]
for all $\{u,v\} \in X$.
The map $\vartheta$ is called a \emph{potential} of $\omega$.
Those well-versed in the parlance of algebraic topology may see that an edge cocycle is a simplicial 1-cocycle and an edge gradient is a simplicial 1-coboundary~\cite[Chapter 3]{hatcher2002algebraic}.
We only need the following standard fact, which can be
proven using a simple argument without the need for any further (cohomological) machinery.

\begin{lemma}\label{lemma:from-cocycle-to-gradient}
  Let $X$ be a simply connected complex. If $\omega$ is an edge cocycle of $X$, then $\omega$ is an edge gradient of $X$.
\end{lemma}
\begin{proof}
  Consider an edge loop $\alpha = (v_0, \ldots, v_k)$. We show that
  \[
    \sum_{i=1}^k \omega(v_{i-1}, v_i) = 0.
  \]
  If $\alpha$ consists of a single vertex, the claim is immediate, since the sum is empty.
  We next show that neither contractions nor their inverses change the sum.
  Consider a contraction $\beta$ of $\alpha$.
  If $\beta$ is obtained by Rule~\labelcref{rule:remove-identical}, then deleting the vertex $v_{i+1}$ does not change the sum as $\omega(v_i, v_{i+1}) = 0$.
  If $\beta$ is obtained using Rule~\labelcref{rule:triangle-contraction}, then some vertex $v_{i+1}$ is removed from the edge loop.
  Since $\omega$ is a cocycle, it satisfies the triangle property \Cref{eq:triangle-property}, and we have $\omega(v_i, v_{i+2}) = \omega(v_i,v_{i+1}) + \omega(v_{i+1}, v_{i+2})$, so the sums do not change.
  Since $X$ is simply connected, the edge loop $\alpha$ can be contracted to a single vertex using contractions and their inverses, and so the sum has to equal zero by induction.

  Next, we define a potential $\vartheta$ for $\omega$, which shows that it is an edge gradient.
  Fix a base vertex $b \in V(X)$.
  Since $X$ is simply connected, it is also path connected. Hence, for each $v \in V(X)$, there exists some edge path $\gamma(v) = (u_0=b, \ldots, u_k=v)$ from $b$ to $v$. Define
  \[
    \vartheta(v) := \sum_{i=1}^k \omega(u_{i-1},u_i).
  \]
  Consider any edge $\{u,v\} \in X$. Let $\alpha = (v_0, \ldots, v_k)$ be the edge loop obtained by going along $\gamma(u)$, going from $u$ to $v$, and returning along the reversal of $\gamma(v)$.  Since $\omega$ is antisymmetric, we get
  \[
    \omega(u,v) + \vartheta(u) - \vartheta(v) =  \sum_{i=1}^k \omega(v_{i-1}, v_i) = 0.
  \]
  This establishes that $\vartheta$ is a potential for $\omega$, so $\omega$ is an edge gradient as claimed.
\end{proof}

\subsection{Synchronous protocol complexes}\label{ssec:topology}

We now explain how executions of synchronous message-passing protocols can be modelled using simplicial complexes, following the axiomatisation of \citet{herlihy2009axiomatic,herlihy2001new}.
The general idea is that we will construct a simplicial complex whose vertices correspond to $r$-round \emph{views} of some process and simplices correspond to views that can occur in the same configuration after $r$ communication rounds.
To define executions of protocols, we will use the notions of pseudospheres, failure operators and round operators. These are simply convenient ways to express how the execution can unfold using the language of combinatorial topology.

\paragraph{Configurations and protocols.}
For each process $i$, a self-stabilising protocol specifies a finite set $S_i$ of states which the process~$i$ can take.
We assume that $S_1, \ldots, S_n$ are pairwise disjoint.
This means that the identity of the process is encoded in its state.
We write $S$ for the union of $S_1,\ldots,S_n$.
Let $\calC$ be the family of sets that contain at most one element from each $S_i$, that is,
\[
  \calC := \left\{ \rho \subseteq S : |\rho \cap S_i| \le 1 \text{ for each process } i \right\}.
\]
Each $\rho \in \calC$ is a \emph{configuration} of the system. Any
$\tau \subseteq \rho$ is a \emph{subconfiguration} of $\rho$.
The set of initial configurations is
\(
  \mathcal{S} := \{ \rho \in \calC : |\rho| = n\}
\),
that is, all configurations where there are initially no crashed processes.
Processes that crash before the start of the protocol can, without loss of generality, be assumed to crash in the first round before sending any messages.

The set $\calC$ also defines all possible \emph{sets} of messages that can be received by a process during the execution of the protocol, as processes always send their current state in each round.
Formally, a protocol defines for each process $i$ a \emph{state transition map}
\(
  \Gamma_i \colon \mathcal{C} \to S_i
\)
and an \emph{output map} $\out_i$ that maps the current state $s \in S_i$  to an output value $\out_i(s)$.
For counting tasks, we use $\Z_q$ as the set of outputs. For pulse synchronisation, the set of outputs will be $\{0,1\}$ whose values indicate whether a pulse is generated or not.

\paragraph{Pseudospheres.}
Let $M$ be a set of values.
A labelling of a simplex $\rho$ with $M$ is a map  $g \colon \rho \to M$. We equivalently treat such a map as the simplex
\[
  \{ (x,g(x)) : x \in \rho \}.
\]
For a simplex $\rho$ and a set $M$ of values, the \emph{pseudosphere}
$M^\rho$ is the set of all labellings of $\rho$ with $M$, where each map $g \in M^\rho$ is interpreted as a simplex. A pseudosphere generates a simplicial complex~\cite{herlihy2009axiomatic}.

\paragraph{The failure and round operators.}
For $k \in \N$, the \emph{$k$-failure operator} is defined as
\[
  \calF_{k}(\rho) := \{ \tau \subseteq \rho : |\rho \setminus \tau| \le k\},
\]
that is, the family of the subsets of $\rho$ with at most $k$ vertices removed.
The idea is that if $\rho$ consists of the \emph{views} of non-crashed processes at the start of a round,
then $\calF_{k}(\rho)$ gives the different sets of views that can influence a process in this round.

A simplex $\rho$ is \emph{between} simplices $\tau$ and $\tau'$ if $\tau \subseteq \rho\subseteq \tau'$.
We write $[\tau:\tau']$ for the set of all simplices between $\tau$ and $\tau'$.
The synchronous \emph{round operator} is a union of pseudospheres given by
\[
  \calR_k(\rho) := \bigcup_{\tau \in \calF_{k}(\rho)}  [\tau:\rho]^\tau.
\]
For a set $A$ of simplices, we write $\calR_k(A) := \bigcup_{\rho \in A} \calR_k(\rho)$, and $\calR^0_k$ is the identity map.
The iterated $(r+1)$-round operator is $\calR^{r+1}_k := \calR_k \circ \calR^r_{k}$ for all $r \ge 0$.
The round operator is used to construct views and configurations that can occur during the execution of a protocol.
The set of $r$-round views in executions with at most $k$ crashes per round is precisely the set of vertices in $\calR^{r}_k(\mathcal{S})$.

\paragraph{Views and executions.}
A 0-round view of process $i$ is an element of $S_i$, as without communication, the process has only information about its own state.
A $t$-round view is a vertex of a simplex in $\calR^t_k(\rho)$ for some $\rho \in \mathcal{C}$ and $k \ge 0$.
Thus, for $t \ge 0$, a $(t+1)$-round view is a pair $(v,M)$ of a $t$-round view $v$ and the set $M$ of $t$-round views
received in round $t+1$.
A view $s \in S_i$ \emph{belongs} to process~$i$, and $(v,M)$ belongs to the same process as $v$.

The state $\mu(v)$ of process $i$ with view $v$ is given recursively by $\mu(s) := s$ for $s \in S$ and $\mu((v,M)) := \Gamma_i(\mu(M))$,
where $\mu(M) := \{ \mu(m) : m \in M \}$.
The output of a view $v$ of process $i$ is $\out(v) := \out_i(\mu(v))$.
For a view $(v,M)$, its \emph{prefix} is $\prefix{(v,M)} := v$, and its \emph{suffix} is
\[
  \suffix{(v,M)} :=
  \begin{cases}
    \mu((v,M)) & \text{if $v$ is a $0$-round view}, \\
    (\suffix{(v)}, \suffix{(M)}) & \text{otherwise},
  \end{cases}
\]
where $\suffix{(M)} = \{ \suffix{(w)} : w \in M\}$. For brevity, we often drop the parentheses when applying the prefix and suffix operations, that is, we write $\prefix{v}$ and $\suffix{v}$.
Note that for any $(t+1)$-round view $v$ of process $i$, both its prefix $\prefix{v}$ and its suffix $\suffix{v}$ are $t$-round views for the same process.

An \emph{execution with at most $k$ crashes per round} is a (possibly infinite) sequence $\rho_0, \rho_1, \ldots$ such that
$\rho_0 \in \mathcal{S}$ and $\rho_{t+1} \in \calR_k(\rho_t)$. The vertices of $\rho_t$ are $t$-round views of the processes that have not crashed by round $t$.
An execution has at most $f$ crashes \emph{in total} if $|\rho_t|\ge n-f$ for all $t \ge 0$. Note that a process can crash only once.
An execution is \emph{crash-free} if it has zero crashes in total.

With a slight abuse of terminology, we also refer to the collection $\rho_t$ of views as a configuration, while technically, $\mu(\rho_t) \in \calC$ is the corresponding configuration of states.
Recall that $S_i$ is the set of states process $i$ can take.
If $\mu(\rho_t) \cap S_i = \emptyset$, then the process $i$ is \emph{crashed} by round $t$.
We say that a process is \emph{non-faulty} in an execution if it never crashes.
Clearly, any finite execution $\rho_0, \ldots, \rho_t$ can be extended to an infinite execution with no further crashes, in which all processes with views in $\rho_t$ remain non-faulty.

\paragraph{The synchronous protocol complex.}
The \emph{$r$-round full-information protocol complex} $R(r,k)$ with \emph{at most} $k$ crashes per round is  the simplicial complex generated by $\calR^r_k(\mathcal{S})$,
that is, the iterated round operator applied to the set of possible initial configurations.
The vertices of $R(r,k)$ correspond to $r$-round views of the processes and simplices to reachable configurations and their subconfigurations in executions with at most $k$ crashes per round. This is the topological object we work with.

Herlihy, Rajsbaum and Tuttle~\cite{herlihy2001new,herlihy2009axiomatic} showed that the round operator preserves high connectivity.
The $r$-round views of self-stabilising protocols correspond to the execution of a non-self-stabilising full-information protocol with input configurations from $\mathcal{S}$.
\Cref{thm:hrt} follows from \cite[Theorem~4.10]{herlihy2009axiomatic} in the same way as it is applied to the synchronous round operator
in the proof of \cite[Theorem~5.10]{herlihy2009axiomatic}.
(Note that \cite{herlihy2009axiomatic} consider $n+1$ processes, whereas we have $n$ processes.)
The main property to verify is that the set $\mathcal{S}$ of initial configurations is an absorbing poset in the sense of
\cite[Definition~3.1]{herlihy2009axiomatic}.
This follows by ordering $\mathcal{S}$ by the lexicographic order $\prec$ on the vectors $(s_1, \ldots, s_n) \in S_1 \times \cdots \times
S_n$.
If $\rho = \{s_1, \ldots, s_n\} \prec \tau = \{s'_1, \ldots, s'_n\}$ and $i$ is the smallest index with $s_i \neq s'_i$, then $\tau_\rho
:= (\tau \setminus \{s'_i\}) \cup \{s_i\}$ satisfies the required conditions.

\paragraph{Bounded-delay model.}
Finally, a subtle technical remark: although the bounded-delay model used for  pulse synchronisation protocols in prior work~\cite{dolev2004self,dolev2014fault,lenzen2019self} usually models local clocks as being real-valued, their numerical values are not accessible to the processes. Instead, the local clocks are used to trigger local state transitions.
We can restrict attention to admissible synchronous executions in which all state transitions occur at discrete time steps and messages always arrive in unit time. Under this restriction, the new state of a process depends only on its current state and the messages received in the round. Hence, any finite-state protocol in this bounded-delay model gives a finite-state synchronous protocol in our setting.

\section{Lower bound for non-trivial counting tasks}\label{sec:non-trivial-lower-bound}

We now prove that any non-trivial counting protocol with $q > 2\sigma$ requires more than $\lfloor f/2 \rfloor$ rounds.

\subsection{Setting up the two synchronous protocol complexes}

Let $\sigma \ge 0$ and $q >\max \{ 2\sigma, 1 \}$.
Suppose that we have a finite-state protocol for non-trivial $\sigma$-approximate $q$-counting that stabilises in $T := \lfloor f / 2 \rfloor$ rounds.
The case $f=0$ is trivial, so we assume $f>0$.
We consider the two synchronous protocol complexes
\[
  X := R(T,2) \qquad \text{ and } \qquad Y := R(T+1,1).
\]
Since $n > f+2 \ge 2(T+1)$ for $X$ and $n > T+2$ for $Y$, \Cref{thm:hrt} implies that the complex $X$ is simply connected and the complex $Y$ is path connected.
Moreover, since the state space of the protocol is finite, both complexes are finite. This will be important.

Every simplex of $X$ or $Y$ is contained in a configuration reached in an execution with at most $2T \le f$ or $T+1 \le f$ crashes, respectively. Since
non-trivial counting only constrains processes that never crash, we can always extend such an execution without further crashes; then every process in the
configuration is non-faulty, and the property \labelcref{def:non-trivial-counting:n1} of non-trivial counting applies to all of them.

We say that $F \colon V(Y) \to V(X)$ is \emph{chromatic} if for any $y \in V(Y)$ the views $y$ and $F(y)$ belong to the same process.
Recall the prefix and suffix maps defined in \Cref{ssec:topology}.
We now establish some useful properties of these maps.

\begin{lemma}\label{lemma:prefix-suffix}
  The prefix $\prefix{} \colon V(Y) \to V(X)$ and suffix $\suffix{} \colon V(Y) \to V(X)$ maps satisfy the following:
  \begin{enumerate}[label=(\alph*)]
    \item The prefix $\prefix{}$ and suffix $\suffix{}$ maps are chromatic simplicial maps from $Y$ to $X$. \label{lemma:prefix-suffix:simplicial}
    \item For any $y \in V(Y)$, the output of a suffix satisfies $\out(\suffix{y}) = \out(y)$. \label{lemma:prefix-suffix:outputs}
    \item If $\rho_0, \ldots, \rho_{T+1}$ is a crash-free execution, then $\prefix{\rho_{T+1}} = \rho_T$, and
      $\suffix{\rho_1}, \ldots, \suffix{\rho_{T+1}}$ is the crash-free execution starting from the configuration $\mu(\rho_1)$. \label{lemma:prefix-suffix:infinite}
  \end{enumerate}
\end{lemma}
\begin{proof}
  Consider a simplex $\tau \in Y$.
  Let $\rho_0, \ldots, \rho_{T+1}$ be a $(T+1)$-round execution with at most one crash per round such that $\tau \subseteq \rho_{T+1}$.
  The prefixes $\prefix{\tau} = \{ \prefix{v} : v \in \tau \}$ are views of the $T$-round execution $\rho_0, \ldots, \rho_T$.
  Hence, these are views of $X$ and $\prefix{\tau} \in X$.

  Now consider the suffix operation.
  If $T=0$, then by definition $\suffix{v} = \mu(v)$ for each 1-round view $v \in \tau$. Thus, $\suffix{\tau}  \subseteq  \mu(\rho_1)$ is a subconfiguration of some initial configuration, and hence, a simplex of $X$. Therefore, assume that $T \ge 1$.
  Let $Q_1$ be the set of processes that crash in the first round and $Q_2$ the set of processes that crash in the second round in the execution.
  Consider the configuration $\mu(\rho_1) = \{ \mu(v) : v \in \rho_1 \}$ consisting of the states of processes that did not crash in the first round. If $Q_1$ is non-empty, it consists of one process $i$. Choose an arbitrary state $s_i \in S_i$ for $i \in Q_1$ and let $\rho_1' := \mu(\rho_1) \cup \{s_i : i \in Q_1\}$ to get an initial configuration where no process has crashed.

  Consider a new execution starting from the configuration $\rho_1'$ in which all processes in $Q_1$ crash immediately before sending any messages.
  The states in $\mu(\rho_1) \subseteq \rho_1'$ are obtained by applying $\mu$ to each 1-round view $v \in \rho_1$, as in this case $\mu$ simply applies the state transition function to the set of messages received by the process with view $v$. Hence, letting a process in $Q_1$ crash before sending any messages in the new execution loses no information, as the information received from $Q_1$ is already reflected in the state $\mu(v)$.
  Now repeat everything as in the second round of the original execution, including crashing the processes in $Q_2$.
  Finally, simulate the remaining rounds $3, \ldots, T+1$ as in the original execution.
  This gives an execution $\rho_1', \ldots, \rho_{T+1}'$ in which at most two processes crash in the first round and at most one process crashes in every subsequent round.
  Thus, $\rho_{T+1}' \in X$ by definition of $X$.

  By construction, the suffix of each $(T+1)$-round view in $\rho_{T+1}$
  is a $T$-round view in $\rho_{T+1}'$.
  Since $X$ is closed under taking subsets, we have that $\{\suffix{v} : v \in \tau \} \in X$.
  Thus, $\prefix{\tau} \in X$ and $\suffix{\tau} \in X$, so the maps are simplicial. Each view is uniquely associated with exactly one process, as states encode the identities of processes. Therefore, the prefix and suffix maps are chromatic.
  This gives \labelcref{lemma:prefix-suffix:simplicial}.
  The claims \labelcref{lemma:prefix-suffix:outputs,lemma:prefix-suffix:infinite} follow easily from the definitions.
\end{proof}

Recall that for a view $v \in V(X)$, we write $\out(v) \in \Z_q$ for the output of a process with view $v$. Next we observe a simple fact that says the outputs in any simplex $\rho \in X$ lie on a $\sigma$-path of $\Z_q$.

\begin{lemma}\label{lemma:outputs-agree}
  Let $\rho \in X$. Then the outputs $\out(\rho)$ lie on a $\sigma$-path of $\Z_q$.
\end{lemma}
\begin{proof}
  A simplex $\rho \in X$ is a subconfiguration that can be reached in some $T$-round execution with at most two crashes per round.
  In total, there are at most $2T \le f$ crashes.
  Let $\phi(i,T)$ be the output of process $i$ for each process whose view is in $\rho$.
  By the property \labelcref{def:non-trivial-counting:n1} of the protocol, there exists a $\sigma$-lift $\lift{\phi}$ from round $T$ onwards.

  By \labelcref{def:sigma-lift:l1}, we have $\phi(i,T) = \lift{\phi}(i,T) \bmod q$.
  By \labelcref{def:sigma-lift:l2}, we have that $|\lift{\phi}(i,T) - \lift{\phi}(j,T)| \le \sigma$.
  Since $\out(v) = \phi(i,T)$ for any process $i$ that has view $v \in \rho$, we get that
  \[
    \out(\rho) \subseteq \{c \bmod q, \ldots, (c+\sigma) \bmod q\}
  \]
  for $c := \min_i \{ \lift{\phi}(i,T) \}$.
  Hence, the outputs lie on a $\sigma$-path of $\Z_q$.
\end{proof}

\begin{figure}[t]
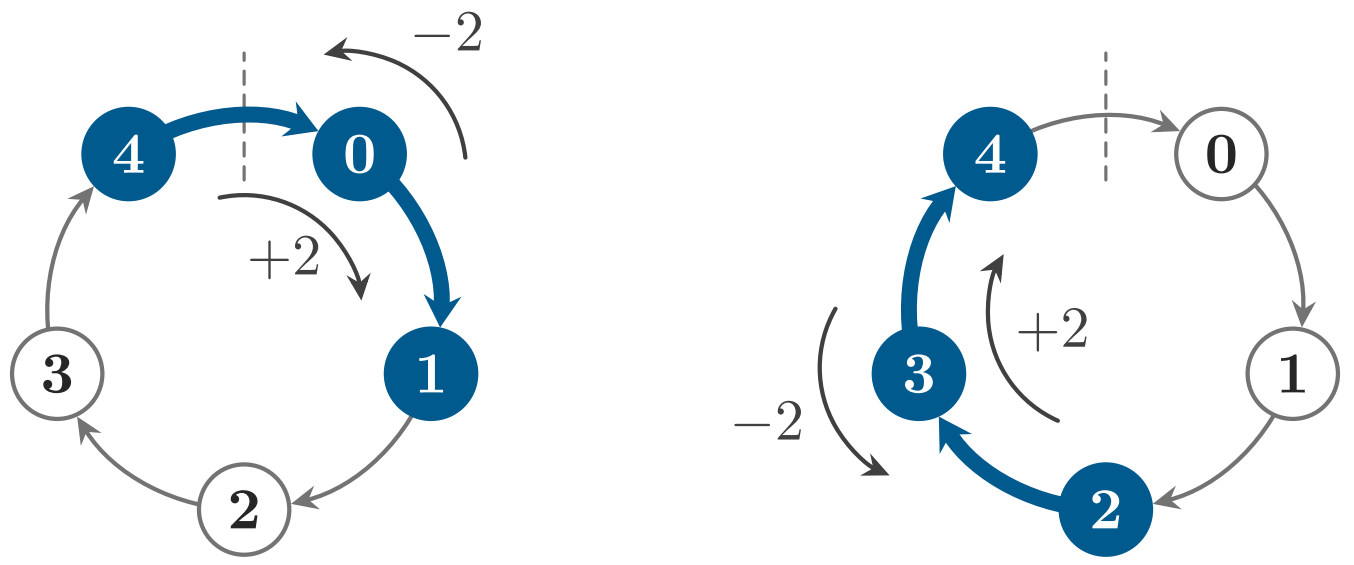

  \centering
  \ClockOffsetsFigure
  \caption{Offsets for $q = 5$ and $\sigma = 2$. Moving clockwise gives a positive offset, moving counter-clockwise a negative offset. The
    dashed line marks the wraparound point between $q-1$ and $0$.
    (a) The path between $a=4$ and $b=1$ crosses the wraparound point. Hence, $d(a,b) = b-a +q = 2$. (b) The path between $a=2$ and $b=4$ does not
  cross the wraparound point, so $d(a,b) = b-a = 2$. }
  \label{fig:offsets-example}
\end{figure}

\subsection{Clock offsets are edge gradients}

Recall that we use the shorthand $[a,b] = \{a, \ldots, b\}$ for integers $a\le b$.
Let $a,b \in \Z_q$ lie on a $\sigma$-path of $\Z_q$.
The \emph{offset} between $a$ and $b$ is the unique integer $d(a,b)$ that satisfies
\begin{enumerate}[label=(O\arabic*),noitemsep]
  \item $d(a,b) \in [-\sigma,\sigma]$, and \label{eq:definition-of-d-first}
  \item $d(a,b) \equiv b-a \pmod q$. \label{eq:definition-of-d}
\end{enumerate}
This is the ``signed'' length of the shortest path from $a$ to $b$ on the $q$-cycle graph, where going clockwise gives a positive sign and going counter-clockwise a negative sign; see \Cref{fig:offsets-example} for an illustration.
This integer is unique because two integers in $[-\sigma,\sigma]$ that are congruent modulo $q$ differ by at most $2\sigma<q$. Put otherwise, on the $q$-cycle, there is only one shortest path between any two values that are distance at most $\sigma$ apart.

For any $\{u,v\} \in X$, we define the \emph{clock offset} of two views $u$ and $v$ as
\begin{equation}\label{eq:clock-offsets}
  D(u,v) := d(\out(u), \out(v)),
\end{equation}
that is, the offset between their outputs. This is well-defined by \Cref{lemma:outputs-agree}, since the outputs lie on a $\sigma$-path.
In the following, we show that the map $D$ is an edge gradient of $X$ with values in $[-\sigma,\sigma]$. We start by showing that $D$ is an edge cocycle.

\begin{lemma}\label{lemma:cocycle}
  The map $D$ is an edge cocycle of the complex $X$.
\end{lemma}
\begin{proof}
  We need to show the triangle property \labelcref{eq:triangle-property}.
  Let $\{u,v,w\} \in X$.
  By \Cref{lemma:outputs-agree}, the outputs $\out(u), \out(v), \out(w)$ lie on some $\sigma$-path $\{c \bmod q,  \ldots, (c+\sigma) \bmod q\}$ of $\Z_q$.
  For each view $x \in \{u,v,w\}$,  choose an integer $\lift{x} \in [c, c+\sigma] \subseteq \Z$ such that
  \(
    \lift{x} \equiv \out(x) \pmod q
  \).

  By definition of $D(x,y)$ in \Cref{eq:clock-offsets}, the clock offset $D(x,y)$ is the only integer in $[-\sigma,\sigma]$ satisfying $D(x,y) \equiv \out(y)-\out(x) \pmod q$.
  Since $\lift{x},\lift{y} \in [c,c+\sigma]$, we have $\lift{y}-\lift{x} \in [-\sigma,\sigma]$,
  and $\lift{y}-\lift{x} \equiv \out(y)-\out(x) \pmod q$.
  Therefore, $D(x,y) = \lift{y} - \lift{x}$.
  This gives the triangle~property:
  \[
    D(u,v) + D(v,w) = \lift{v} - \lift{u} + \lift{w} - \lift{v} = \lift{w} - \lift{u} = D(u,w). \qedhere
  \]
\end{proof}

We now use the fact that on a simply connected complex $X$, an edge cocycle is an edge gradient. This allows us to find a potential $\ell \colon V(X) \to \Z$ that lifts the output values from $\Z_q$ to $\Z$.

\begin{lemma}\label{lemma:lifting}
  There exists a map $\ell \colon V(X) \to \Z$ such that
  \begin{enumerate}[label=(\alph*)]
    \item $\ell(v) - \ell(u) = D(u,v)$ for all $\{u,v\} \in E(X)$, \label{lemma:lifting:ell-is-potential}
    \item $\out(u) = \ell(u) \bmod q$  for all $u \in V(X)$. \label{lemma:lifting:ell-congruent-to-out}
  \end{enumerate}
\end{lemma}
\begin{proof}
  By \Cref{lemma:cocycle}, the map $D$ is an edge cocycle of $X$. Since $X$ is simply connected, \Cref{lemma:from-cocycle-to-gradient} implies that $D$ is an edge gradient, that is, there exists a map $\vartheta \colon V(X) \to \Z$ such that $D(u,v) = \vartheta(v) - \vartheta(u)$.
  Fix an arbitrary vertex $b \in V(X)$. For each $u \in V(X)$,  define
  \[
    \ell(u) := \vartheta(u) - \vartheta(b) + \out(b).
  \]
  Now for any edge $\{u,v\} \in E(X)$, we have that
  \begin{equation}\label{eq:ell-identity}
    \ell(v) - \ell(u) = \vartheta(v) - \vartheta(u) = D(u,v).
  \end{equation}
  This gives claim \labelcref{lemma:lifting:ell-is-potential}.

  We now show claim \labelcref{lemma:lifting:ell-congruent-to-out}. Fix $u \in V(X)$. Since $X$ is path connected, there exists some edge path $\alpha = (u_0 = b, \ldots, u_k = u)$ from $b$ to $u$. By \Cref{eq:ell-identity}, we have
  \begin{equation*}
    \ell(u) - \ell(b) = \sum_{i=1}^k \big( \ell(u_i) - \ell(u_{i-1}) \big) = \sum_{i=1}^k D(u_{i-1}, u_i).
  \end{equation*}
  By definition of $D$, we have that $D(u_{i-1}, u_i) \equiv \out(u_i) - \out(u_{i-1}) \pmod q$. Hence, we get
  \begin{equation*}
    \sum_{i=1}^k D(u_{i-1}, u_i) \equiv \sum_{i=1}^k \big( \out(u_i) - \out(u_{i-1}) \big) \pmod q.
  \end{equation*}
  The sum on the right-hand side telescopes, so putting the above together yields
  \begin{alignat*}{3}
    &      &\ell(u) - \ell(b) &\equiv \out(u) - \out(b) &\quad &\pmod q \\
    \iff{} &&\ell(u)          &\equiv \out(u)           &      &\pmod q,
  \end{alignat*}
  where the equivalence follows from $\ell(b) = \out(b)$.  This gives claim \labelcref{lemma:lifting:ell-congruent-to-out}.
\end{proof}

\subsection{Clock offset consistency across rounds}

\begin{figure}[t]
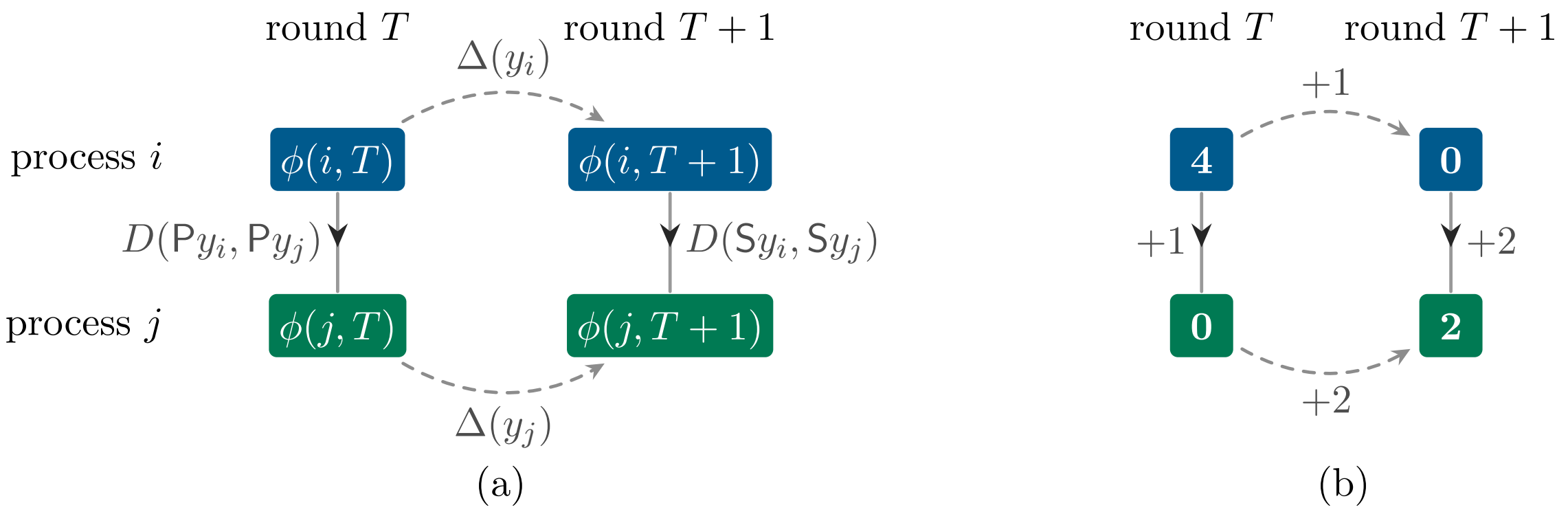

  \centering
  \OffsetConsistencyFigure
  \caption{Offset consistency. (a) The outputs of process $i$ (view $y_i$) and process $j$ (view $y_j$) in rounds $T$ and $T+1$. The dashed arrows
    are the clock advancements and the solid arrows the clock offsets from $i$ to $j$. (b) An example with $q = 5$ and $\sigma = 2$, with vertices
    coloured by process as in \Cref{fig:gradient-illustration}. Here, process $i$ outputs $4$ in round $T$ and advances its clock by one to output $0$
    in round $T+1$. Process $j$ outputs $0$ in round $T$ and then advances its clock by two to output $2$ in round~$T+1$.
    Note the difference with \Cref{fig:gradient-illustration-c}, where approximate counting guaranteed $\Delta(\cdot)=1$.
  }
  \label{fig:offset-square}
\end{figure}

By \Cref{lemma:prefix-suffix}, the prefix $\prefix{}$ and suffix $\suffix{}$ maps take vertices from $Y$ to $X$.
For any $(T+1)$-round view $y \in V(Y)$, the \emph{advancement} is the difference
\begin{equation}\label{eq:delta-definition}
  \Delta(y) :=  \bigl( \out(\suffix{y}) - \out(\prefix{y})  \bigr) \bmod q,
\end{equation}
which
measures  how much the process with view $y$ advanced its clock between rounds $T$ and $T+1$.
We show that this difference must be consistent with the clock offsets in the complex $X$; see \Cref{fig:offset-square}.

\begin{lemma}\label{lemma:offset-consistency}
  For any $\{y_i,y_j\} \in E(Y)$, we have
  \[
    D(\suffix{y_i}, \suffix{y_j}) + \Delta(y_i) = D(\prefix{y_i}, \prefix{y_j}) + \Delta(y_j).
  \]
\end{lemma}
\begin{proof}
  By definition of $Y$, the vertices $y_i$ and $y_j$ of the edge correspond to the $(T+1)$-round views of non-crashed processes $i$ and $j$ in some execution with at most $f$ crashes. Fix this execution.
  Let $\phi(i, \cdot)$ and $\phi(j, \cdot)$ be the outputs of these processes in this execution; we only consider the outputs in rounds $T$ and $T+1$.
  By \Cref{lemma:prefix-suffix}\labelcref{lemma:prefix-suffix:simplicial}, $\prefix{y_i}$ and $\suffix{y_i}$ are vertices of the complex $X$. Moreover,
  \begin{equation}\label{eq:out-identity}
    \out(\prefix{y_i}) = \phi(i,T) \qquad \text{ and } \qquad \out(\suffix{y_i}) = \phi(i,T+1),
  \end{equation}
  where the first equality is by definition and the second holds by \Cref{lemma:prefix-suffix}\labelcref{lemma:prefix-suffix:outputs}.
  The same holds analogously for $y_j$.

  Since the protocol solves non-trivial counting, by property~\labelcref{def:non-trivial-counting:n1} in \Cref{def:non-trivial-counting}, there exists a $\sigma$-lift $\lift{\phi}$ of the outputs of this execution from round $T$ onwards. Fix $\lift{\phi}$ and let $\delta := \lift{\phi}(j,T) - \lift{\phi}(i,T)$. Observe now that
  \begin{itemize}[noitemsep]
    \item By \labelcref{def:sigma-lift:l1}, we have $\delta \equiv \phi(j,T) - \phi(i,T) \pmod q$.
    \item By \labelcref{def:sigma-lift:l2}, we have $\delta \in [-\sigma, \sigma]$.
  \end{itemize}
  Since $\delta$ satisfies the above two properties, by definition of $d$, we get that $\delta = d(\phi(i,T), \phi(j,T))$. Hence,
  \[
    D(\prefix{y_i}, \prefix{y_j}) =  d(\phi(i,T), \phi(j,T)) = \delta =
    \lift{\phi}(j,T) - \lift{\phi}(i,T).
  \]
  Analogously, we also get the identity
  \begin{align*}
    D(\suffix{y_i}, \suffix{y_j}) &= \lift{\phi}(j,T+1) - \lift{\phi}(i,T+1).
  \end{align*}
  Consider $\Delta(y_i)$ and let $\eta := \lift{\phi}(i,T+1) - \lift{\phi}(i,T)$. Now
  \begin{itemize}[noitemsep]
    \item By \labelcref{def:sigma-lift:l1}, we have $\eta \equiv \phi(i,T+1) - \phi(i,T) \pmod q$.
    \item By \labelcref{def:sigma-lift:l3}, we have $\eta \in [0,q-1]$.
  \end{itemize}
  Hence, we get that $\eta = \left( \phi(i,T+1) - \phi(i,T) \right) \bmod q$.
  Now from \Cref{eq:delta-definition} and \Cref{eq:out-identity}, we get
  \begin{align*}
    \Delta(y_i) &= \bigl( \out(\suffix{y_i}) - \out(\prefix{y_i})  \bigr) \bmod q \\
    &= \left(\phi(i,T+1) - \phi(i,T) \right) \bmod q \\
    &= \lift{\phi}(i,T+1) - \lift{\phi}(i,T).
  \end{align*}
  The identity for $\Delta(y_j)$ is analogous.
  Combining the above, we get
  \begin{align*}
    D(\suffix{y_i}, \suffix{y_j}) - D(\prefix{y_i}, \prefix{y_j})
    &= \bigl[\lift{\phi}(j, T+1) - \lift{\phi}(i, T+1)\bigr] - \bigl[\lift{\phi}(j, T) - \lift{\phi}(i, T)\bigr] \\
    &= \bigl[\lift{\phi}(j, T+1) - \lift{\phi}(j, T)\bigr] - \bigl[\lift{\phi}(i, T+1) - \lift{\phi}(i, T)\bigr] \\
    &= \Delta(y_j) - \Delta(y_i).
  \end{align*}
  Rearranging gives the claim.
\end{proof}

\begin{lemma}\label{lemma:uniform-advance}
  Let $\ell$ be the map from \Cref{lemma:lifting}.
  There exists $c \in \Z$ such that for all $y \in V(Y)$,
  \[
    \ell(\suffix{y}) - \ell(\prefix{y}) = \Delta(y) + qc.
  \]
\end{lemma}
\begin{proof}
  Define the map $\epsilon \colon V(Y) \to \Z$ as
  \[
    \epsilon(y) := \ell(\suffix{y}) - \ell(\prefix{y}) - \Delta(y).
  \]
  By \Cref{lemma:lifting}\labelcref{lemma:lifting:ell-congruent-to-out} and \Cref{eq:delta-definition}, we get that
  \(
    \epsilon(y) \equiv 0 \pmod q
  \).
  We now show that $\epsilon$ is constant on $V(Y)$.
  Let $\{y_i, y_j\} \in E(Y)$. Since $\prefix{}$ and $\suffix{}$ are chromatic simplicial maps from $Y$ to $X$, we get that
  $\{ \prefix{y_i}, \prefix{y_j}\}, \{ \suffix{y_i}, \suffix{y_j} \} \in E(X)$. Rearranging and invoking \Cref{lemma:lifting}\labelcref{lemma:lifting:ell-is-potential} gives
  \begin{align*}
    \epsilon(y_i) - \epsilon(y_j) &= \Bigl[\ell(\suffix{y_i}) - \ell(\prefix{y_i}) - \Delta(y_i)  \Bigr]- \Bigl[\ell(\suffix{y_j}) - \ell(\prefix{y_j}) - \Delta(y_j)  \Bigr] \\
    &= D(\suffix{y_j}, \suffix{y_i}) - D(\prefix{y_j}, \prefix{y_i}) + \Delta(y_j) - \Delta(y_i) = 0,
  \end{align*}
  where the last equality follows from \Cref{lemma:offset-consistency}. Since the complex $Y$ is path connected, we get that $\epsilon$ is constant on $V(Y)$. Since also $\epsilon \equiv 0 \pmod q$, the claim follows.
\end{proof}

We are finally ready to prove \Cref{thm:non-trivial-lower-bound}.

\nontriviallb*
\begin{proof}
  Suppose that such a protocol exists.
  Let $\ell \colon V(X) \to \Z$ be the map given by \Cref{lemma:lifting} and $c \in \Z$ be the constant from \Cref{lemma:uniform-advance}.
  By \labelcref{def:non-trivial-counting:n2}, there exists a crash-free execution $(\rho_t)_{t \ge 0}$ where some process $i$ changes its output infinitely often. For all $t \ge 0$, let $x_t \in V(X)$ be the $T$-round view of  $i$ in the crash-free execution started from configuration~$\mu(\rho_t)$, and
  $y_t \in V(Y)$ be the $(T+1)$-round view of  $i$ in the crash-free execution started from configuration~$\mu(\rho_t)$.

  Observe that $\prefix{y_t} = x_t$ and $\suffix{y_t} = x_{t+1}$ by \Cref{lemma:prefix-suffix}\labelcref{lemma:prefix-suffix:infinite}. Define $\Delta_t := \Delta(y_t)$ and note that $\Delta_t \in \{0,1, \ldots, q-1\}$.
  From \Cref{lemma:uniform-advance}, we get that
  \(
    \ell(x_{t+1}) - \ell(x_t) = \Delta_t + qc
  \).
  Rewriting yields
  \begin{equation}\label{eq:punchline}
    \ell(x_{t}) = \ell(x_0) + \sum_{r=0}^{t-1} \big( \Delta_r + qc \big).
  \end{equation}
  By property \labelcref{def:non-trivial-counting:n2}, we get that $\Delta_t \ge 1$ for infinitely many $t$, because the output $\out(x_t)$ of $i$ changes infinitely often in the execution $(\rho_t)_{t \ge 0}$.
  Now consider two cases:
  \begin{itemize}[noitemsep]
    \item
      If $c \ge 0$, then the sum on the right-hand side tends to $\infty$ as $t \to \infty$.
    \item If $c \le -1$, then $qc + \Delta_r \le -1$, and the sum on the right-hand side tends to $-\infty$ as $t \to \infty$.
  \end{itemize}
  By \Cref{eq:punchline}, this means that $\ell \colon V(X) \to \Z$ takes infinitely many values.
  However, the complex $X$ is finite, which is a contradiction.
\end{proof}

\section{Upper bound for approximate counting}\label{sec:upper-bound}

In this section, for the sake of completeness, we show an upper bound of
$\lceil f/2 \rceil + O(\log q)$ rounds
for 1-approximate $q$-counting for $q \ge 4$.
For $q=4$, we get stabilisation in $\lceil f/2 \rceil + 1$ rounds, which matches exactly the lower bound given in \Cref{corollary:apx} for any even $f > 0$.

One attempt at obtaining the bound is to use a direct reduction to cycle agreement via the pipelining construction of \citet{daliot2005self} and a terminating $T$-round cycle agreement protocol~\cite{alistarh2023wait}. However, this would give a protocol that stabilises in $2T$ rounds, which is not tight, as $T > \lfloor f/2 \rfloor$ for any deterministic cycle agreement protocol.
Instead, we construct essentially a self-stabilising version of cycle agreement, where outputs advance by one modulo $q$ each round.
However, unlike in terminating cycle agreement, we cannot first run 2-set agreement, and then switch to approximate agreement on a path (because this would require agreement on time).

\subsection{The approximate counting protocol}

Let $\phi(i,t) \in \Z_q$ be the state and output of process $i$ at the end of round $t$. Since we are considering a self-stabilising protocol, the values $\phi(i,0)$ are arbitrary.
We now treat $\Z_q$ as the vertices of a $q$-cycle graph.
For a set $M \subseteq \Z_q$, choose a canonical shortest path $a_0, \ldots, a_k$, where $a_{h} = a_0 + h \bmod q$, that spans the vertices in $M$. Let $s(M)$ be the length of this path and $c(M) = a_0 + \lfloor k/2 \rfloor \bmod q$ be the vertex in the center of this path. The values in $M$ lie on an edge if $s(M) \le 1$.

\paragraph{The protocol.}
For each  $t \ge 0$, let $M(t)$ be the set consisting of the outputs $\phi(i,t)$ for each process $i$ that does not crash by the end of round $t$.
In each round $t+1$, every non-crashed process $i$ sends its previous output $\phi(i,t)$ to all processes, including itself, and receives a set $M(i,t) \subseteq M(t)$.
Then
\begin{enumerate}[noitemsep]
  \item if $M(i,t)$ does not lie on an edge, set $\phi(i,t+1) :=  c(M(i,t)) + 1 \bmod q$, and
  \item otherwise, set $\phi(i,t+1) := \phi(i,t) + 1 \bmod q$.
\end{enumerate}
That is, if the received values do not lie on an edge, then go to the center of the shortest path spanning the vertices received. Otherwise, the values look fine, and the process increments its value by one modulo $q$. The last rule ensures that stabilisation persists once reached.
We will call the first rule the ``center rule'' and the latter the ``edge rule''.

\paragraph{The analysis.}
The analysis is close to the standard cycle agreement analysis~\cite{alistarh2023wait} adapted to the self-stabilising setting. Within $\lceil f/2 \rceil+1$ rounds, there will be a round with at most one crash. Once this happens, there can be only two distinct sets of received messages among the non-crashed processes, and hence, at the start of the next round there will be at most \emph{three} values in play. However, these values will be close to one another.
We say that $M \subseteq \Z_q$ is \emph{good} if $s(M) \le \lceil q/4 \rceil$.

\begin{lemma}\label{lemma:good-round}
  Let $q \ge 4$.
  There exists a round $t \le \lceil f/2 \rceil + 1$ such that $M(t)$ is good.
\end{lemma}
\begin{proof}
  Consider the first round $t_0$ with at most one crash.
  Let $\chi \in \{0,1\}$ be the number of crashes in round~$t_0$. As every earlier round has
  at least two crashes, we have $2(t_0-1) + \chi \le f$, and hence $t_0 + \chi \le \lceil f/2 \rceil + 1$. We show that $M(t)$ is good for some $t \le t_0 + \chi$.

  Since $t_0$ has at most one crash, there are two message sets $A,B$ such that $A \subseteq B \subseteq M(t_0-1)$ received by the processes. The value of any process that does not crash in the round is in $A$.
  If $t_0$ is crash-free, then $A=B$.
  If $t_0$ had one crash, then $B = A \cup \{v\}$.
  Consider the following three cases:
  \begin{enumerate}
    \item Suppose $B$ lies on an edge. Then $A$ lies on an edge and all processes apply the edge rule, and $M(t_0)$ is good.
    \item Suppose $A$ lies on an edge and $B$ does not.
      Let $b_0, \ldots, b_k$ be vertices on the fixed shortest path spanning $B$.
      Now $A \subseteq \{b_0,b_1\}$ or $A \subseteq \{b_{k-1}, b_k\}$.
      Of the two paths from $v$ to $A$, the longer has length at least $\lceil (q-1)/2 \rceil$.
      Hence, $k \le q - \lceil (q-1)/2 \rceil \le \lceil q/2 \rceil$.
      The processes that receive message set $B$ choose their output to be $b_{\lfloor k/2 \rfloor+1}$ by the center rule.  By the edge rule, the outputs of processes that receive the message set $A$ will be either in $\{b_1,b_2  \}$ or $\{b_k, b_{k+1}\}$, where $b_{k+1} \neq b_{k-1}$ is the other vertex adjacent to $b_k$. Thus,
      the distance between the new output values is at most $\lceil k/2 \rceil \le \lceil q/4 \rceil$. Hence, $M(t_0)$ is good.

    \item Suppose $A$ does not lie on an edge. Then $B$ does not lie on an edge, and all processes apply the center rule.
      If $M(t_0)$ has exactly one value, then it is good.
      Suppose  there are two distinct values in $M(t_0)=\{a,b\}$. This means that round $t_0$ had a crash, so $\chi=1$.
      Any two processes $i$ and $j$ that do not crash in round $t_0+1$ will always receive each other's values.
      Hence, the only two possible message sets are $\{a\}$ and $\{a,b\}$, without loss of generality.

      If $\{a,b\}$ lies on an edge, all processes apply the edge rule, and so the new outputs lie on an edge.
      Otherwise, let $x$ be the vertex in the center of the canonical shortest path from $a$ to $b$. Then the outputs are contained in $\{a+1 \bmod q, x+1 \bmod q\}$. Since $a$ and $b$ are at most distance $d \le \lceil q/2 \rceil$ apart, $x$ is at distance  at most $\lceil d/2 \rceil$ from $a$. Hence, the outputs are within distance at most $\lceil q/4 \rceil$ from one another, so $M(t_0+1)$ is good. \qedhere
  \end{enumerate}
\end{proof}

Once we reach a round with a good $M(t)$, we are now essentially running approximate agreement on a path and the argument is standard. More precisely, we have the following property.

\begin{lemma}\label{lemma:halving}
  Let $t \ge 0$ and $q \ge 4$.
  If $M(t)$ is good, then
  \[
    s(M(t+1)) \le \left\lceil \frac{s(M(t))}{2} \right\rceil.
  \]
  Moreover, $M(t+1)$ is good as well.
\end{lemma}
\begin{proof}
  Let $I = \{a_0, \ldots, a_k \}$ be vertices on the fixed shortest path spanning $M(t)$.
  By definition,
  vertices in $M(t)$ lie on this path.
  Since $M(t)$ is good, we have $k \le \lceil q/4 \rceil < q/2$.
  Therefore, any shortest path between any two vertices in $I$ also lies in $I$.

  Consider two processes $i$ and $j$ that do not crash in round $t+1$. Let $x_i,x_j \in M(t)$ be their values at the end of round $t$. Let $y_i, y_j \in M(t+1)$ be their values at the end of round $t+1$. The message sets received by $i$ and $j$ both contain $x_i,x_j$, as $i$ and $j$ did not crash by round $t+1$. Now we have three cases:
  \begin{enumerate}
    \item If both processes used the edge rule, then $x_i,x_j$ lie on an edge, and so do $y_i$ and $y_j$.
    \item Suppose that process $i$ used the center rule and process $j$ did not.
      Since $y_i - 1 \bmod q$ is in the center of the shortest path spanning $M(i,t)$, which contains $x_j$, we get that $y_i$ is within distance at most $\lceil k/2 \rceil$ from $y_j$.
    \item Suppose that both processes used the center rule. Then the spanning subpaths used by $i$ and $j$ intersect and contain $x_i$, and hence, $y_i-1$ and $y_j-1$ are at most $\lceil k/2 \rceil$ apart, as they are both centers of their respective subpaths.
  \end{enumerate}
  Hence, $s(M(t+1)) \le \lceil k/2 \rceil \le k$ and $M(t+1)$ is good.
\end{proof}

\begin{theorem}\label{thm:apx-upper-bound}
  Let $q \ge 4$.
  There is a 1-approximate $q$-counting protocol that stabilises by round
  \[
    \lceil f/2 \rceil + \lceil \log_2 q \rceil -1.
  \]
\end{theorem}
\begin{proof}
  By \Cref{lemma:good-round}, there is some round $t \le \lceil f/2\rceil +1$ when $M(t)$ is good. Then apply \Cref{lemma:halving} $K := \lceil \log_2 q \rceil - 2$ times.
  Since $s(M(t)) \le \lceil q/4 \rceil$, for $T=t+K$, we get
  \[
    s(M(T)) \le \left \lceil \frac{q}{2^{K+2}} \right\rceil \le 1,
  \]
  so $M(T)$ lies on an edge and remains on an edge, and all processes apply the edge rule for all future rounds by induction.
  Hence, the protocol stabilises in $T$ rounds, which gives the claim.
\end{proof}

\section{Conclusions}\label{sec:conclusions}

In this work, we proved the first non-trivial and asymptotically tight lower bound of $\Omega(f)$ rounds for approximate clock synchronisation
tasks that guarantee both transient and non-transient fault tolerance. The lower bound holds in a strong model: processes have perfect local clocks
without drift, communication is synchronous, and the only non-transient faults are crashes. Existing upper bounds work even when the non-transient faults are Byzantine, providing very high levels of resilience.

Approximate clock synchronisation tasks that require only Byzantine fault tolerance or only self-stabilisation, but not both, have been known to be considerably easier to solve. Our results show that achieving a higher level of resilience by combining both types of fault tolerance necessarily comes with a high price in worst-case stabilisation time -- at least for deterministic algorithms.

To the best of our knowledge, this is the first use of the topological framework of distributed computing to prove lower bounds for
self-stabilising tasks.
For accessibility, we kept the exposition combinatorial, but in topological terms, the proof is a covering-space argument: as simply connected
complexes have ``vanishing first cohomology'' (i.e., every 1-cocycle is a 1-coboundary), we can lift the output map from the protocol complex
to the universal cover $\Z$ of the $q$-cycle $\Z_q$.
The prefix and suffix maps then show that the lifted outputs of any process in two consecutive rounds differ by its clock increment plus a
multiple of $q$ that is the same for all processes and executions.
In a crash-free execution in which some process changes its output infinitely often, the lifted outputs of this process diverge, contradicting
the finiteness of the protocol complex.

This leads to a few further questions.
First, are there network-dependent lower bounds for clock synchronisation in general communication networks, analogous to those for $k$-set
agreement~\cite{fraigniaud2025lower}?
Second, what is the broader role of cohomology in the topology of distributed computing~\cite{felber2025brief,hylton2023toward}?
Finally, are there topological characterisations of the solvability of self-stabilising tasks, analogous to those known for terminating
tasks~\cite{herlihy2013distributed}?

\subsection*{Acknowledgements}

The author is grateful to Faith Ellen for many useful discussions and her insightful observations, which helped clarify the main ideas and the approach taken in this work.
The author also thanks Yaroslav Verbitsky for his helpful feedback on an earlier draft of the manuscript.

\subsection*{AI methodology}

The author used GPT-6 and Claude Opus 5.5 to assist with finding and developing the lower bound construction, drawing illustrations, and improving the presentation.
The tools materially affected drafting and streamlining technical parts of \Cref{sec:overview,sec:preliminaries,sec:non-trivial-lower-bound,sec:upper-bound}. The text has been written by the author, but copyediting has been done using AI tools.
The idea of reducing pulse synchronisation to relaxations of approximate counting is by the author.
In particular, there was a weaker gluing-based lower bound for a restricted class of crash-tolerant protocols that predates the use of AI tools.

The author first posed the question of whether set agreement and cycle agreement lower bounds extend to self-stabilising
variants under crash faults. Then the author posed the same question for the approximate counting task. For the latter question, GPT-6 eventually supplied the first prefix--suffix construction, which compared overlapping execution windows, using the two synchronous full-information protocol complexes obtained from~\cite{herlihy2009axiomatic}. The original construction used a more involved covering-space argument.
Claude Opus 5.5 was used to verify and simplify the construction.
Further simplifications, and the non-trivial counting generalisation used for the pulse synchronisation reduction, were developed interactively with both AI tools.
The author verified the correctness and originality of all content, including references, and takes full responsibility for any~errors.

\printbibliography

\end{document}